**Systematic study of high performance GeSn photodiodes with thick absorber for SWIR and extended SWIR detection.**

*Quang Minh Thai, Rajesh Kumar, Abdulla Said Ali, Justin Rudie, Steven Akwabli, Yunsheng Qiu, Mourad Benamara, Hryhorii Stanchu, Kushal Dahal, Xuehuan Ma, Sudip Acharya, Chun-Chieh Chang, Gregory T. Forcherio, Bruce Claflin, Wei Du and Shui-Qing Yu\**

Q. M. Thai, R. Kumar, A. S. Ali, J. Rudie, S. Akwabli, Y. Qiu, S. Acharya, W. Du, S. -Q. Yu
Department of Electrical Engineering and Computer Science, University of Arkansas, Fayetteville, Arkansas 72701, USA
E-mail: syu@uark.edu

A. S. Ali, J. Rudie, K. Dahal, X. Ma, W. Du, S. -Q. Yu
Material Science and Engineering Program, University of Arkansas, Fayetteville, Arkansas 72701, USA

M. Benamara, H. Stanchu, W. Du, S. -Q. Yu
Institute for Nanoscience and Engineering, University of Arkansas, Fayetteville, Arkansas 72701, USA

C. -C. Chang
Center for Integrated Nanotechnologies, Los Alamos National Laboratory, Los Alamos, New Mexico, 87545, USA

G. T. Forcherio
Naval Surface Warfare Center, Electro-Optics Tech. Div. 300 Highway 361, Crane, Indiana 47522, USA

B. Claflin
Air Force Research Laboratory, Sensors Directorate, Wright-Patterson Air Force Base, Dayton, Ohio 45433, USA

**WILEY-VCH**

**Funding**:

The authors acknowledge funding support from Office of Naval Research (ONR) (Grant No. N00014-24-1-2651). J. Rudie acknowledges the funding support from the DoW SMART Scholarship.

**Abstract** Germanium-tin (GeSn) photodiodes potentiate a viable solution to integrate SWIR and extended SWIR detection technology into CMOS processing line. However, challenges in the growth of thick, high quality GeSn limit the device absorber thickness, making it impossible to ascertain the performance limit of GeSn photodiodes. An in-depth understanding of their device physics and a clear optimization pathway towards commercial-grade devices remain elusive. This work presents a systematic empirical study of GeSn photodiodes with thick absorber (2 to 8% Sn content, up to 2630 nm thick), showing high responsivity up to 0.59 A.W$^{-1}$ at 1.55 μm and 0.43 A.W$^{-1}$ at 2 μm wavelengths, low dark current density down to 2 x 10$^{-2}$ A.cm$^{-2}$, and high detection cutoff wavelengths up to 2.1 and 2.5 μm at 5% and 8% Sn, respectively. Using specific doping design (P-i-N and N-i-P), an in-depth analysis is presented on the impact of junction position, p-type background carrier concentration, bulk/ surface defects and photocarrier diffusion length - on photodetection performance. Different optimization strategies for GeSn photodiodes, in particular at high Sn content, are proposed.

## 1. Introduction

Photodetection capability in short-wave infrared (SWIR, 0.9 – 1.7 μm) and extended SWIR (e-SWIR, up to 2.5 μm) range is crucial for many applications, spanning LiDAR, autonomous driving, environmental monitoring, to computer vision, thanks to its ability to provide position target discrimination in extreme conditions such as dust, fog or smoke while maintaining "eye-safe" compatibility. Ge and InGaAs photodiodes are among the most common detectors in SWIR/ e-SWIR range, with the latter capable of providing excellent detection beyond 2 μm, up to 2.6 μm thanks to its tunable band gap [1]. However, integrating InGaAs technology into Silicon (Si) CMOS fabrication line remains challenging, due to difficult InGaAs monolithic growth on Si and potential cross-contamination of III-V elements as accidental dopants in Si. Successful demonstration of germanium-tin (GeSn) monolithic growth on Si substrate for laser and detector in SWIR-MIR range offers a novel, alternative material approach to SWIR/e-SWIR photodiode [2–50], with a clear pathway towards monolithic, CMOS-integrated focal plane array (FPA) design - as previously demonstrated in a commercial Ge-on-Si CMOS imager [51], while aiming towards detection performance level comparable to InGaAs device.

However, achieving high performance GeSn photodiodes in SWIR and in particular e-SWIR range, with simultaneously low dark current density (InGaAs reference - 500 μm device: $1.5 \times 10^{-5}$ A.cm$^{-2}$ to $1.5 \times 10^{-3}$ A.cm$^{-2}$, for cutoff wavelength increased from 1.9 to 2.6 μm [1]), high responsivity (InGaAs reference: around 1 and 1.2 A.W$^{-1}$ at 1.55 μm and 2 μm wavelengths [1]) and high detection wavelength range remains a difficult task. Due to a large mismatch between GeSn and Ge lattice parameters, GeSn epitaxial growth is sensitive to defect and dislocation from plastic relaxation as the layer thickness increases. The majority of GeSn photodiodes reported so far were designed with a thin GeSn absorber (often between 100-500 nm): while such approach limits the junction width (i.e. depletion width) and therefore can reduce the dark current density, it also limits the device responsivity and detection range, the latter due to band gap increase from high residual compressive strain in GeSn layer. As a result, thin GeSn absorber provides little information regarding the material limit of GeSn photodiode performance, and it is not suitable in the long term for further device optimization towards commercial-grade quality. It is thus crucial to conduct a systematic study for GeSn photodiodes with thick absorber (between 1 and 5 μm thick) at different Sn contents, with the junction positioned in high crystalline quality region to provide a fair comparison with the state of the art of common IR photodiodes, extract an in-depth understanding of its device physics, identify current technological bottlenecks and finally devise efficient optimization strategies.

In this article, we report results for photodiodes with thick GeSn absorber (950 to 2630 nm thick), with Sn content from 2% Sn to 8% Sn, showing low dark current density (down to $1.6 \times 10^{-2}$ and $2.4 \times 10^{-2}$ A.cm$^{-2}$ at 2% and 5% Sn), high responsivity in SWIR/ e-SWIR range (0.37 - 0.59 A.W$^{-1}$ at 1.55 μm, 0.29 - 0.43 A.W$^{-1}$ at 2 μm) and extended detection cutoff wavelength (up to 2.49 μm at 8% Sn). A simultaneous presence of low dark current density ($2.4 \times 10^{-2}$ A.cm$^{-2}$ at 300 K), high responsivity (0.39 A.W$^{-1}$ at 1.55 μm, 0.33 A.W$^{-1}$ at 2 μm, 300 K) and high detection cutoff wavelength (2.08 μm at 300 K) was achieved in 5% Sn photodiode, thanks to a specific doping profile (P-i-N, p+ contact layer on top) which confined the junction in high crystalline quality GeSn layer, while distancing and thus protecting it from surface defects. Finally, with results from two doping profile designs (P-i-N and N-i-P, the latter with n+ contact layer on top), the impact of different factors (junction position, junction width, photocarrier diffusion length, bulk/surface defects) on device dark current density and responsivity was analyzed at different Sn contents.

## 2. Results and discussion

## 2.1 Design overview (P-i-N/ N-i-P) – Material growth - Experimental methodology

As p-type background carrier concentration was previously reported for Ge and GeSn epitaxial growth [44,52–54], with values in $10^{15}$ – $10^{17}$ cm$^{-3}$ range, the junction width can be significantly reduced and became shorter than the absorber thickness. Therefore, the doping profile had a major impact on the junction position and the device performance as a result (**Figures 2**): in P-i-N structure, the junction was positioned near the bottom interface (n+ doped), far from the top surface (p+ doped), while the junction in N-i-P structure was positioned near the top surface (n+ doped) instead. The impact from surface defect on the device performance, as well as the main photocarrier transport mechanism (drift/ diffusion) were thus expected to be different between the P-i-N and N-i-P design. A summary of structure designs and performance results of all P-i-N and N-i-P GeSn photodiodes in this work was provided in **Table 1**.

***Figure 1.*** *Schematic for a) P-i-N and b) N-i-P GeSn photodiodes in this work, with the electric field and photocarrier type – diffusion ($e^-_{diff}$) and drift ($e^-_{drift}$) – indicated. The intrinsic absorber layer was assumed to be unintentionally p-doped. Device was illuminated from the top surface (front illumination). Bottom contact was placed on either the Ge buffer or the GeSn spontaneous relaxation enhanced (SRE) layer.*

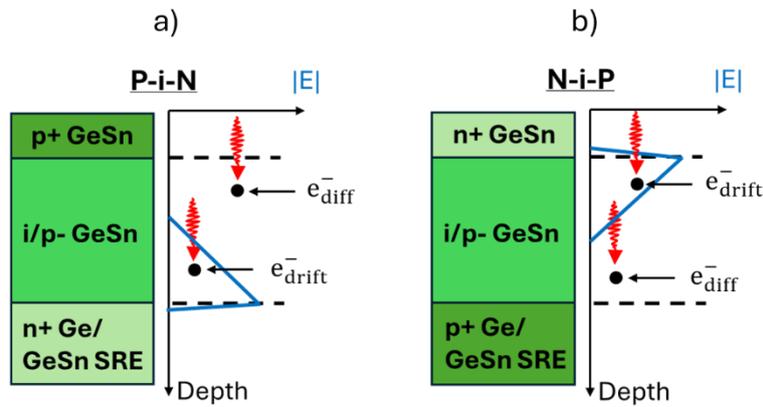

***Table 1.*** *Summary table for structure designs and performance results of all GeSn photodiodes presented in this work. $d_{tot}$, $d$, $J_d$, $J_d/d$, $R$ represented total GeSn thickness, GeSn absorber thickness (undoped or with p-type background carrier concentration), dark current density (taken at -1 V), dark current density normalized by the absorber thickness, maximum responsivity (at 1.55 μm and 2 μm), respectively. The cutoff wavelength was defined at 50% of the peak responsivity.*

| Structure | Sn content (%) | $d_{tot}$ (nm) | $d$ (nm) | $J_d$ (A.cm$^{-2}$) | $J_d/d$ (A.cm$^{-3}$) | $R$ 1.55 µm (A.W$^{-1}$) | $R$ 2 µm (A.W$^{-1}$) | Cutoff wavelength (µm) |
|---|---|---|---|---|---|---|---|---|
| | 2 | 2300 | 2180 | $1.6 \times 10^{-2}$ | $7.3 \times 10^{1}$ | 0.51 | -- | 1.74 |
| P-i-N | 5 | 2600 | 1310 | $2.4 \times 10^{-2}$ | $1.8 \times 10^{2}$ | 0.39 | 0.33 | 2.08 |
| | 8 | 1800 | 950 | $4.7 \times 10^{-1}$ | $4.9 \times 10^{3}$ | 0.37 | 0.41 | 2.49 |
| | 2 | 2700 | 2630 | $1.7 \times 10^{-1}$ | $6.5 \times 10^{2}$ | 0.59 | -- | 1.77 |
| N-i-P | 5 | 1600 | 1545 | $7.4 \times 10^{-2}$ | $4.8 \times 10^{2}$ | 0.48 | 0.29 | 2.08 |
| | 8 | 1800 | 1760 | $6.6 \times 10^{-1}$ | $3.7 \times 10^{3}$ | 0.46 | 0.43 | 2.47 |

All GeSn structures in this work were grown on top of 1200 nm-thick relaxed Ge buffer, using an ASM Epsilon 2000 reduced pressure chemical vapor deposition (RPCVD) reactor, with germane (GeH$_4$) and tin tetrachloride (SnCl$_4$) as precursors and H$_2$ as carrier gas. Detailed CVD GeSn growth method was reported elsewhere [55]. p+ doping and n+ doping for contact layers were performed using diborane (B$_2$H$_6$) and phosphane (PH$_3$) gas, with doping concentration between 5 x 10$^{17}$ and 5 x 10$^{18}$ cm$^{-3}$. X-ray diffraction data with (224) reciprocal space mapping (RSM) were provided in **Supplementary Information, Figure SI1**.

Full characterizations of GeSn photodiode performance were performed from 77 K to 300 K, with current-voltage (I-V), capacitance-voltage (C-V) and responsivity spectral response characteristics. I-V characteristics were measured using Keysight B2911B source measure unit (SMU), C-V measurements were performed using a Keithley 590 CV Analyzer, and responsivity spectral responses were measured using a ThermoFisher Nicolet 8700 FTIR, equipped with an IR source. The IR source was calibrated using an InGaAs photodiode (Thorlabs FD10D with cutoff wavelength at 2.6 µm). Responsivities at 1.55 µm and 2 µm were measured using continuous 1.55 µm laser from BKtel Photonics and continuous tunable laser from IPG Photonics respectively, with optical power measured via a thermal sensor (Thorlabs S401C). Device was illuminated in front-side configuration (i.e. from the top surface).

Responsivities were calculated either directly from dark/ light I-V characteristics, or from noise-rejection lock-in amplifier setup, with both results cross-checked against each other.

## 2.2 P-i-N photodiodes results

Schematics of 2%, 5% and 8% Sn P-i-N photodiodes, alongside their SIMS profiles were shown in **Figure 2**. For 5% Sn and 8% Sn structures, as spontaneous relaxation enhanced (SRE) GeSn defective layer was formed during the growth [56,57] (see **Supplementary Information, Figure SI2** for TEM images), the bottom GeSn layers (1200 nm and 800 nm thick for 5% Sn and 8% Sn respectively) was n+ doped to confine the depletion region in the upper GeSn layers of better crystalline quality, aiming to reduce impact from bulk defect (threading dislocation) on the device dark current density. For 2% Sn structure, as the layer thickness remained close to the critical epitaxial thickness [57], n+ doping was solely performed on the Ge buffer.

**Figure 2.** *Schematic of a) 2% Sn, b) 5% Sn and c) 8% Sn P-i-N photodiodes. Yellow regions indicated gold contacts. SIMS profiles of d) 2% Sn, e) 5% Sn and f) 8% Sn P-i-N photodiodes showed doping profiles (P, B) and Ge, Sn concentration profiles.*

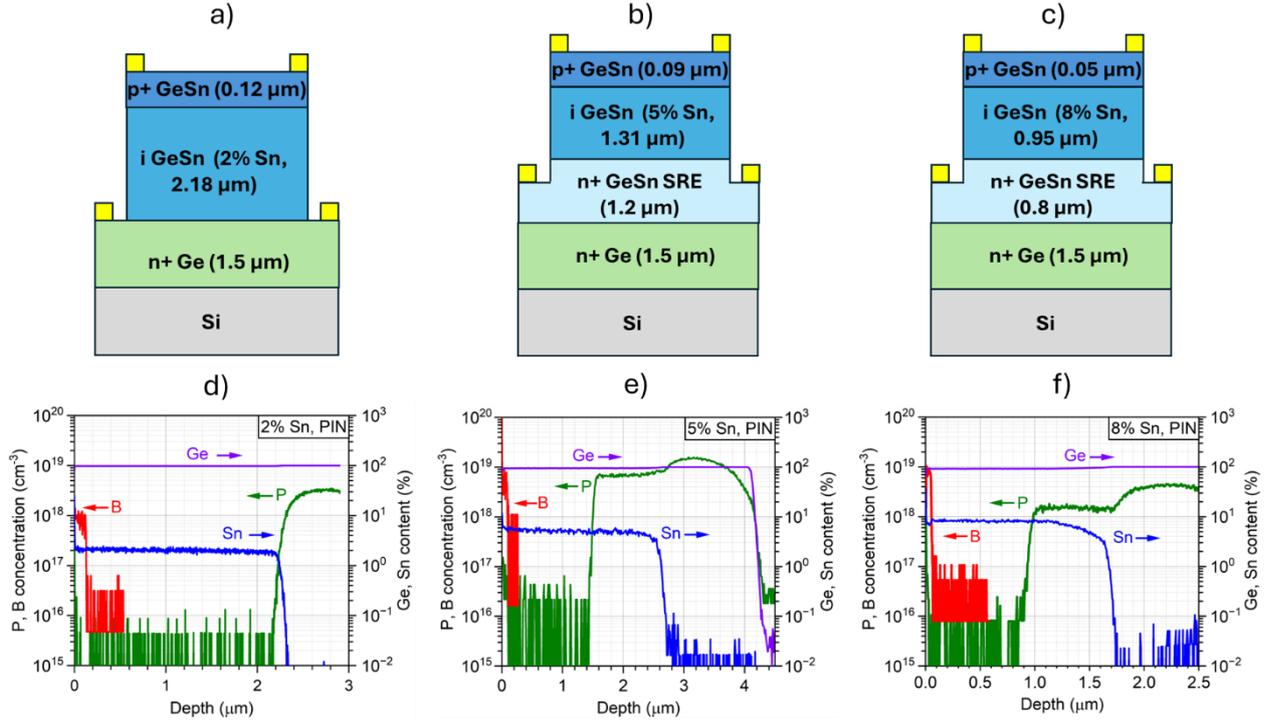

**Figures 3a-c** showed dark and light I-V characteristics at 300 K for all three samples, under 100 µW optical power at 1.55 and 2 µm wavelength. On 2% Sn and 5% Sn P-i-N photodiodes, clear distinction between dark/ light I-V characteristics and low dark current density (1.6 x 10⁻² and 2.4 x 10⁻² A.cm⁻² at -1 V, for 2% Sn and 5% Sn respectively) was observed. On 8% Sn P-i-N photodiode, photocurrent remained strong but higher dark current density (4.7 x 10⁻¹ A.cm⁻²) made it harder to distinguish between dark/ light I-V. By plotting dark current density as function of $\frac{1}{k_BT}$ (**Figure 3d**), we extracted the activation energy $E_a$ using Arrhenius equation:

$$J_d = J_0 e^{-\frac{E_a}{kT}} \tag{1}$$

with $J_0$, $E_a$ fitting parameters, $k$, $T$ the Boltzmann constant and temperature. Between 200 K and 300 K, high $E_a$ values of 0.232 eV and 0.174 eV were extracted for 2% Sn and 5% Sn - relatively close to their mid-bandgap values, while extremely low $E_a$ value of 0.030 eV was observed on 8% Sn. It indicated a transition of dark current mechanism from Shockley-Read Hall generation-recombination (SRH GR) to trap-assisted tunneling (TAT) as Sn content increases, suggesting significantly higher threading dislocation density in 8% Sn absorber. At low temperature regime (below 200 K), all $E_a$ values were extremely low, indicating TAT as the main dark current mechanism in this regime regardless of Sn content. TAT-dominated dark current at low temperature was observed in other heteroepitaxial photodiodes in SWIR/ e-SWIR range and beyond, like Ge-on-Si photodiode [58] or type-II superlattice (T2SL) InAs/GaSb photodiode [59,60]. It is also worth noting that for 8% Sn, dark current density barely changed with temperature, since TAT dark current density only weakly depended on the temperature via the trap energy $E_g - E_t$, while SRH GR dark current had stronger dependence to the temperature via the intrinsic carrier concentration $n_i \propto e^{-\frac{E_g}{2kT}}$. Detailed dark current analysis to quantify contribution from SRH GR and TAT dark current density can be found in **Supplementary Information, section SI2**.

**Figure 3.** *Dark and light I-V characteristics for a) 2% Sn b) 5% Sn and c) 8% Sn P-i-N photodiode at 300 K. At 2% Sn, only light IV at 1.55 μm was shown, since its cutoff wavelength is well below 2 μm. c) Dark current density of GeSn P-i-N photodiodes at -1 V as function of $\frac{1}{kT}$, for temperatures from 77 K to 300 K, with Arrhenius fit to extract activation energies $E_a$*

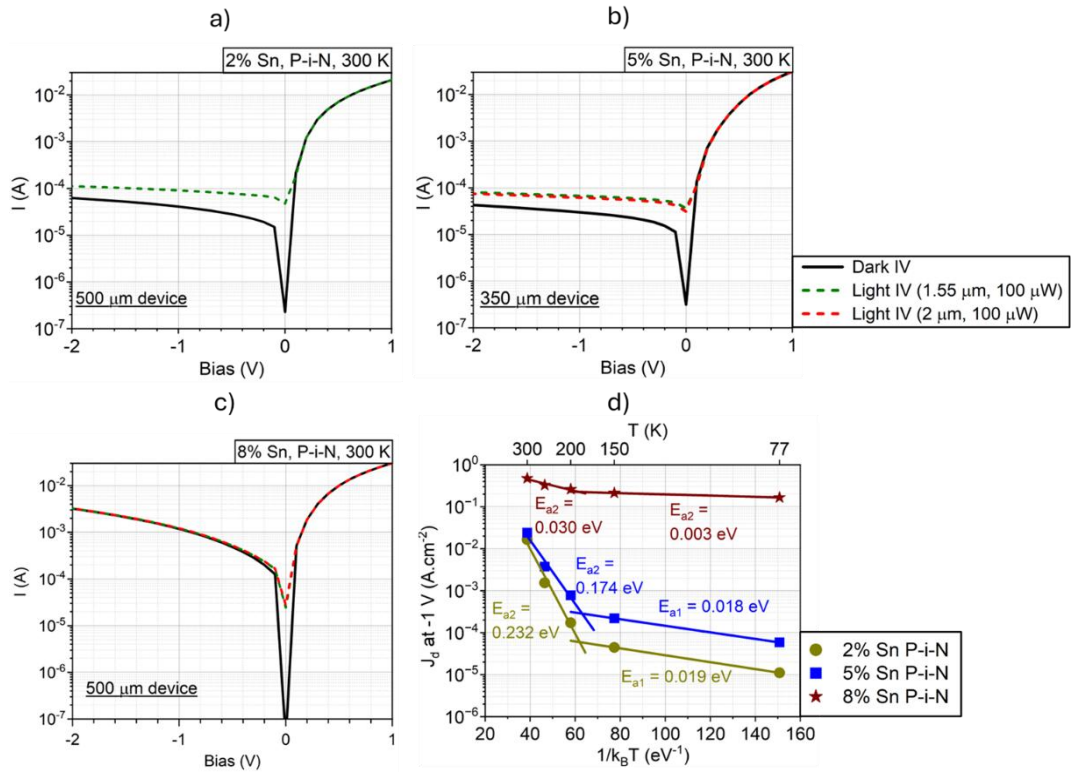

1.55 and 2 µm responsivities for all P-i-N photodiodes at 300 K were plotted in **Figure 4a** for reverse bias between 0 and -2 V. Stable responsivities were observed for 2% Sn and 5% Sn P-i-N photodiodes, while a brief increase of responsivities at low bias was observed at 8% Sn before stabilizing. 2 µm responsivity only existed in 5% Sn and 8% Sn, with their cutoff wavelengths beyond 2 µm at 300 K (**Figures 4b,c**). At 5% Sn, 2 µm responsivity was slightly lower than that of 8% Sn (0.33 A.W⁻¹ and 0.41 A.W⁻¹, respectively), with a significant increase of responsivity with temperature. This was due to a cutoff wavelength below 2 µm at low temperature (1.87 µm at 77 K), which then increased up to 2.08 µm at 300 K, higher but remained very close to 2 µm.

**Figure 4.** *a) 1.55 and 2 µm responsivity as function of reverse bias at 300 K, for 2%, 5% and 8% Sn P-i-N photodiodes. b) 2 µm maximum responsivity for 5 % Sn and 8% Sn P-i-N photodiodes as function of temperatures. c) Responsivity FTIR spectral response of P-i-N photodiodes at 77 K (top) and 300 K bottom. Sharp spectral response cutoff at 2.55 µm for 8%*

*Sn at 300 K was due to the cutoff of the reference InGaAs photodiode. It should be noted that slight difference between FTIR responsivity and that directly measured with 1.55 and 2 μm laser (plots a), b)) might be observed.*

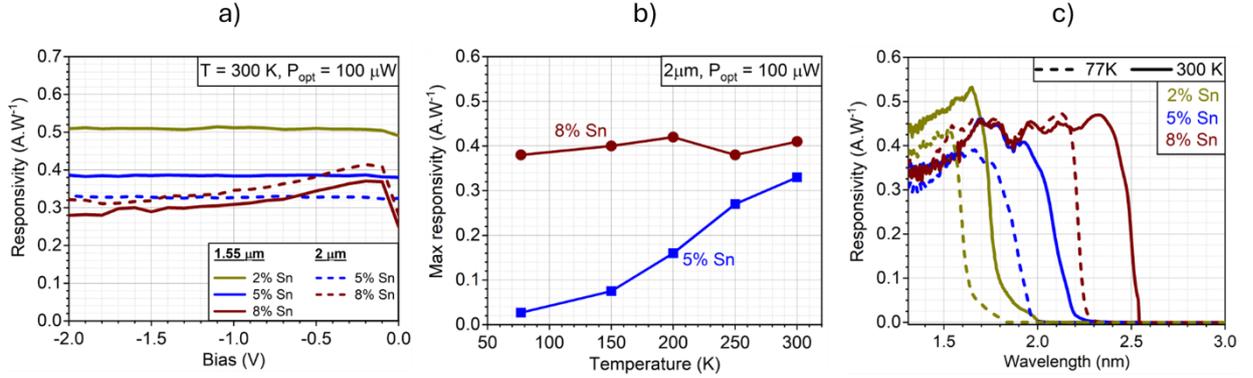

To further interpret these results, we conducted C-V measurement to extract the junction width $W$ and p-type background carrier concentration profile $N_A(W)$ in the absorber, based on the following formulae (assuming that junction width in the contact layer was negligible) [61,62]:

$$W = \frac{K_s \, \varepsilon_0 \, A}{C} \qquad (2)$$

$$N_A = -\frac{2}{q K_s \varepsilon_0 A^2 \frac{\partial \left(\frac{1}{C^2}\right)}{\partial V}} \qquad (3)$$

with $C, K_s, A$ the measured capacitance, dielectric constant and device area. Results were shown in **Figure 5** for 2% and 5% Sn photodiodes, showing a decrease of the depletion width as Sn content increased – from 350-1100 nm (2% Sn) to 250-600 nm (5% Sn) - alongside an increase of background carrier concentration – from $2 \times 10^{15}$ - $4 \times 10^{15}$ cm$^{-3}$ (2 % Sn) to $4 \times 10^{15}$ – $1 \times 10^{16}$ cm$^{-3}$ (5% Sn). Attempted C-V measurement for 8% Sn (*not shown here*) returned unphysical results, with junction width larger than GeSn total thickness for reason remaining yet unknown. C-V measurement for 8% Sn N-i-P sample was shown instead later in **Section 2.3** for comparison. Here, one observed that the junction width was significantly shorter than the absorber thickness in both cases, as previously mentioned in **Section 2.1**: the junction of 2% and 5% Sn P-i-N photodiodes was thus buried deep inside the absorber, distanced from the top surface. Therefore, the impact from surface defects on the dark current density was expected to be limited in these cases. In addition, as significant number of photons were absorbed on GeSn region near top surface, especially at shorter wavelength like 1.55 µm, diffusion was likely the main photocarrier transport mechanism in GeSn P-i-N photodiodes under front illumination. Stable responsivity regardless of reverse bias, as previously shown in Figure 4a, indicated an absence of field enhancement effect and further supported this hypothesis. By subtracting the lowest depletion width at 0 V from the absorber thickness, and knowing that photocarrier can only be absorbed within one diffusion length from the junction, one can grossly estimate that the diffusion length of minority carrier (here, electron) was at least 1-2 µm in 2% and 5% Sn P-i-N structures, indicating a high crystalline quality in the absorber. In contrast, slight field enhancement effect can be observed in 8% Sn P-i-N photodiode, which might be the indication of both shorter junction width and diffusion length in this sample. A detailed discussion of this phenomenon and what it means for higher Sn content photodiode was provided later in **Section 2.4**.

**Figure 5.** *Plots of a) capacitance as function of reverse bias, b) junction width as function of reverse bias and c) p-doping profile as function of junction width for 2 and 5% Sn P-i-N. All measurement and data analysis were conducted at 300 K.*

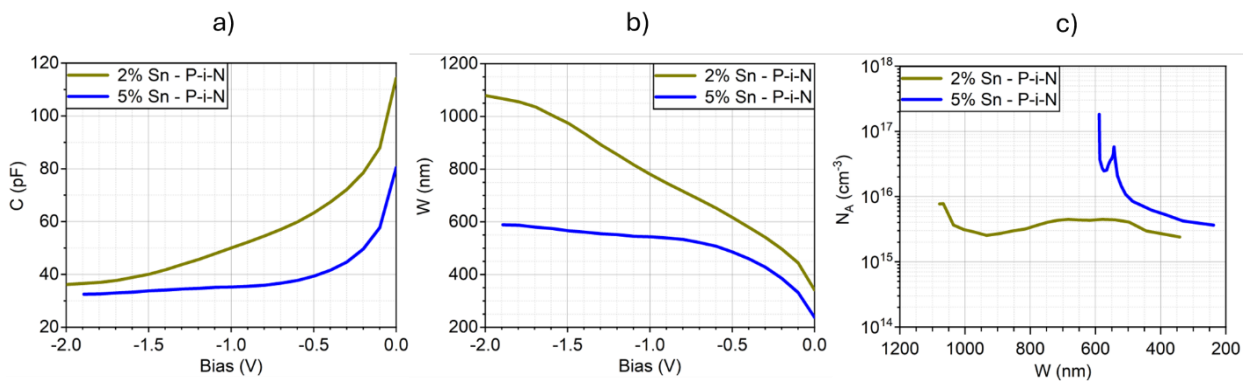

Compared to other results reported so far for GeSn photodiodes in similar Sn content range, to the best of our knowledge, dark current density, responsivity and cutoff wavelength of P-i-N photodiode in this work ranked among the best values (**Figure 6**). Dark current density values (**Figure 6a - top**) of 2 and 5% Sn photodiode in this work were among the lowest values reported so far for GeSn photodiodes: for a better indication, as the GeSn layer here was significantly thicker than that reported in other work, plot of dark current density normalized by absorber thickness was also provided (**Figure 6a - bottom**), showing once again lowest values for 2% and 5% Sn P-i-N photodiode. At 8% Sn, even when the dark current density significantly increases, normalized value showed that it remained close to best results reported so far, which was acceptable when taking account of its high responsivity and cutoff wavelength altogether. While a few other works showed higher responsivity at 1.55 µm (**Figure 6b - top**), 2 µm responsivity at 5 and 8% Sn (**Figure 6b - bottom**) and their cutoff wavelengths (**Figure 6c**) clearly showed a significant improvement compared to other results in similar Sn content range, which was attributed to a thicker GeSn absorber and consequently a higher relaxation degree. Biasing the device beyond -2 V, as reported in Refs. [17,27,30], can further enhance the responsivity, but at the cost of higher dark current and thus higher detector noise. One can notice that 5% Sn P-i-N photodiode in this work simultaneously satisfied three important criteria for e-SWIR photodiode: low dark current density, high responsivity with a detection cutoff wavelength beyond 2 µm.

**Figure 6.** *Comparison of P-i-N photodiodes in this work with other GeSn photodiodes results reported in literature. a) Dark current density (top) and normalized dark current density by absorber thickness (bottom). b) 1.55 (top) and 2 µm (bottom) maximum responsivity. c) Detection cutoff wavelength (50% of peak intensity). All data were taken at 300 K.*

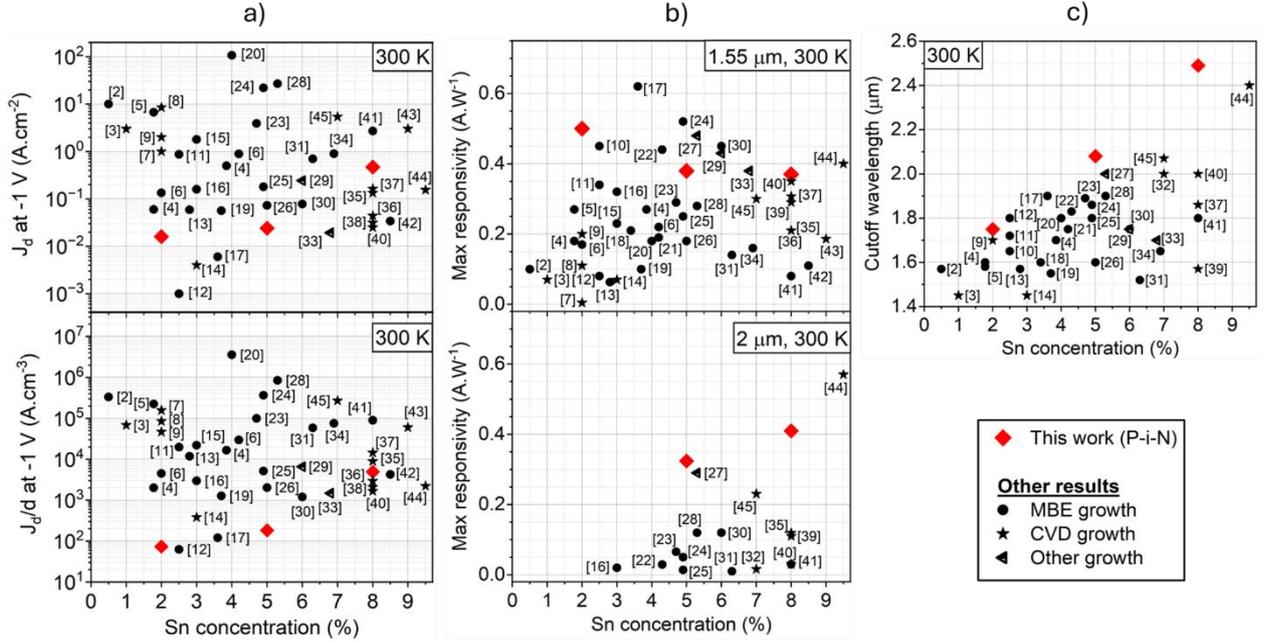

Specific detectivities ($D^*$) of all P-i-N photodiodes in photovoltaic mode (PV - 0 V bias) were plotted in **Figures 7 a-c,** showing comparable or better performance at 300 K compared to some common IR detector like InAs photodiode or PbSe photoconductor, which was further improved when cooling down to 250 K (i.e. typical temperature achieved with single-stage thermoelectric cooler) or to cryogenic temperature (77 K). In PV mode, with thermal noise as the main contribution, $D^*$ can be calculated as [63]:

$$D^* \approx \frac{R\sqrt{A}}{i_{thermal}} = \frac{R\sqrt{R_0 A}}{\sqrt{4kT}} \qquad (4)$$

with $i_{thermal}$, $R$, $A$, $T$, $R_0$ the thermal noise current, responsivity, device area, temperature and the device resistance at 0 V bias. **Figure 7d** showed $R_0 A$ values as function of temperature for P-i-N GeSn photodiodes in this work, showing comparable or higher values than InAs, InSb, MCT photodiodes, explaining high $D^*$ value of GeSn photodiodes. It should be noted that at 300 K, $R_0 A$ values of GeSn P-i-N photodiodes remained at least two orders of magnitude below those of commercial Ge and InGaAs detectors, showcasing the challenge of maintaining good material quality in heteroepitaxy of layers with high lattice mismatch. However, as the growth strategy for high quality GeSn layer followed quite a similar approach which was used for InGaAs e-SWIR photodiode (i.e. with closely lattice-matched substrate like InP and graded buffer layer of InAlAs or InAsP [64–71]), further optimization of the GeSn growth recipe via annealing process or extremely thick graded buffer layer can significantly enhance its material quality and thus increase $R_0A$ and $D^*$ values altogether. For example, the relative difference of lattice parameters between fully relaxed 5% and 8% Sn GeSn and Ge was 0.7% and 1.2% (0.6% and 1.0% if taking account of the tensile strain in Ge buffer), respectively. These values were indeed comparable with those of InGaAs absorber with cutoff wavelength up to 2 – 2.6 µm, where Indium concentration varied between 70 – 85%, corresponded to 1.2 - 2.2% of lattice parameters relative difference with InP substrate.

**Figure 7.** *Plots of D\* in PV mode of a) 2% Sn, b) 5% Sn and c) 8% Sn P-i-N photodiode at 77 K, 250 K and 300 K compared to other commercial IR detectors, either photodiode in PV mode or photoconductor (PC). d) Plot of $R_0A$ as a function of temperature for both samples, compared to other commercial IR detectors. List of commercial detector references used in this figure: InGaAs (G17190, G17192, G17193, Hamamatsu); InSb (P5968, Hamamatsu); InAs (P10030-01 (300 K), P7163 (77 K), Hamamatsu); T2SL (P15409-901, Hamamatsu); MCT (J19TE4:5-3VN-R250U, Teledyne Judson); Ge (J16-18A-R500U-HS, Teledyne Judson). PbS and PbSe PC data were taken from Ref. [63]*

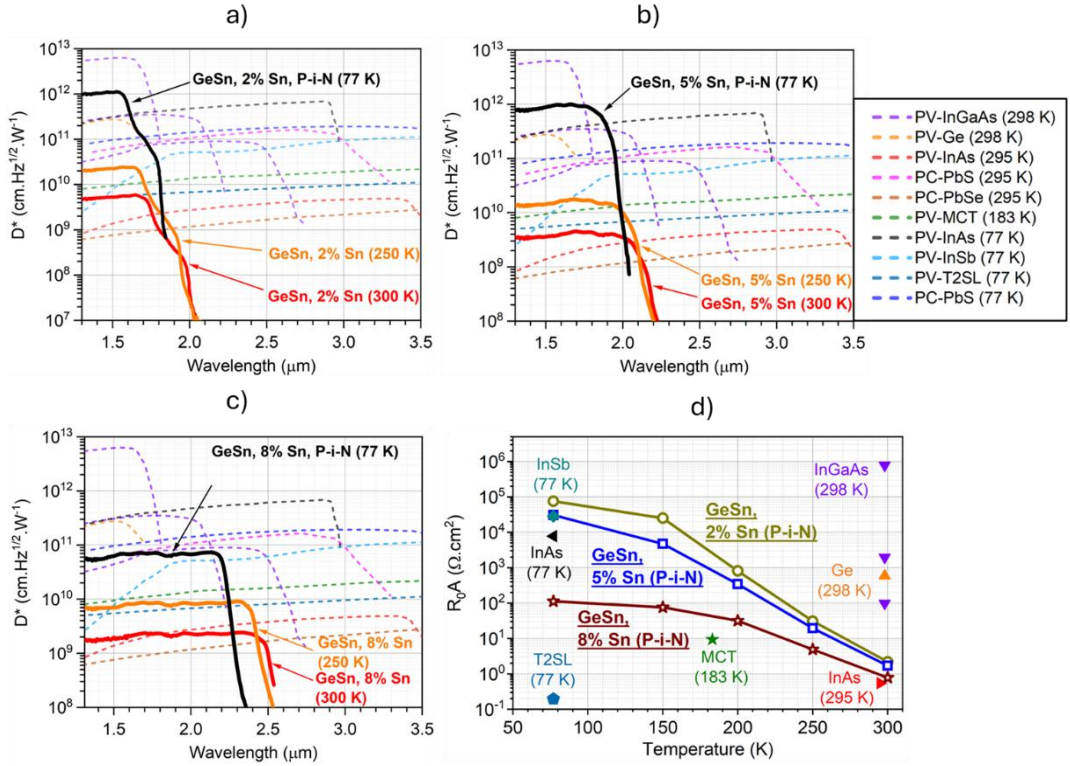

### 2.3 N-i-P photodiodes results

While P-i-N design showed promising results as seen from section 2.2, one can notice that deeply buried junction in such case might not be the best configuration for photocarrier collection. For IR photons heavily absorbed near the top layer, while they can diffuse to the depletion region, they might also recombine via surface defects in proximity and therefore resulted in a loss in responsivity. Detailed calculation of drift and diffusion photocurrent also predicted that at similar absorber thickness, diffusion current was lower than the drift current by a factor of $\frac{\alpha L_e}{1+\alpha L_e}$, with $\alpha$ and $L_e$ the light absorption coefficient and electron diffusion length, respectively [61]. For such reason, a set of GeSn N-i-P photodiodes with Sn content from 2 to 8% Sn was designed and grown, with detailed structure shown in **Figure 8 a-c.** With n+ top contact layer and an unintentional p-doped absorber, the junction was expected to be positioned close to the top surface and the main photocarrier transport mechanism switched from diffusion to drift, aiming for an increase of device responsivity. As the junction width was shorter than the absorber thickness, it did not include the SRE layer in 5 and 8% Sn structure; therefore, we only performed p+ doping on the Ge layer for bottom contact layer.

**Figure 8.** *Schematic of a) 2% Sn, b) 5% Sn and c) 8% Sn N-i-P photodiodes. d), e), f) Dark and light IV (1.55 and 2 μm) at 300 K for all N-i-P photodiodes.*

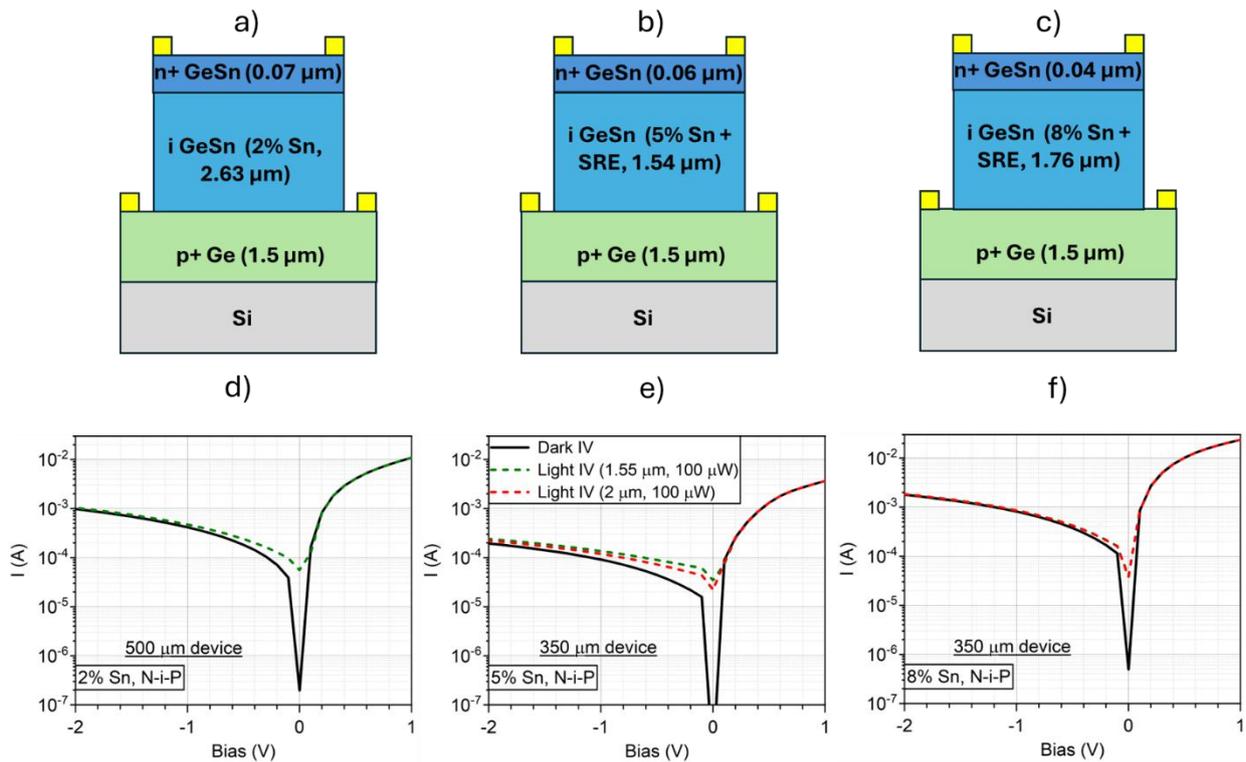

From dark and light IV plots of N-i-P photodiodes shown in **Figure 8d-f**, one can observe that dark and light IV were less distinguishable in this case compared to previous P-i-N results. It was due to an increase in the dark current density (**Figure 9a**) on N-i-P samples for all Sn content, most notable at 2% and 5% Sn. We attributed such phenomenon to the amplified leakage from surface defects, due to junction position close to the top surface. At 8% Sn, P-i-N and N-i-P dark current densities were relatively similar: as speculated from section 2.2, here, bulk defects (threading dislocation) remained the main contribution to the dark current, which made contribution from surface defects less pronounced.

On the other hand, 1.55 µm responsivity showed a consistent increase of 0.08 – 0.09 A.W$^{-1}$ on all N-i-P photodiodes compared to their P-i-N counterpart (**Figure 9b**), which we attributed to a switch from diffusion to drift transport mechanism for photocarrier. While one might argue that slightly thicker GeSn absorber of N-i-P photodiodes compared to P-i-N might also be an important contribution to this phenomenon, it should be noted that the increase in responsivity was constant across all samples, regardless of the thickness variation. In 5% and 8% Sn N-i-P photodiodes in particular, the absorber region also included the SRE layer, which was highly defective and therefore made photocarrier generated there prone to recombination before reaching the junction (i.e. low diffusion length), significantly counteracted the benefit of thicker absorber on the responsivity, if it occurs. 2 µm responsivity remained very similar between N-i-P and P-i-N photodiodes, which we attributed to longer penetration depth at this wavelength, making 2 µm responsivity less sensitive to the change in junction position. Finally, C-V measurement on 8% Sn N-i-P photodiode revealed a low junction width between 250 – 550 nm and an p-type background carrier concentration above 6 x 10$^{15}$ cm$^{-3}$, comparable to previous results of 5% Sn P-i-N photodiode (**Figure 9c,d**). Junction width and doping profile for 2% and 5% Sn N-i-P photodiodes (see **Supplementary Information, Figure SI9**) were comparable with their P-i-N counterparts.

**Figure 9.** *a) P-i-N and N-i-P normalized dark current density. Theoretical diffusion limit for dark current density from Ref.* [72] *was shown in dashed pink line. b)1.55 and 2 µm maximum responsivity for P-i-N and N-i-P photodiodes. Theoretical maximum responsivity (without anti-reflective coating) was shown in dashed black and red lines. Theoretical 2µm maximum responsivity at 8% Sn was shown in cross-circle symbol, taken account of GeSn thickness (with and without SRE layer) and assuming that absorption coefficient is 10000 cm$^{-1}$* [73]. *c), d) Junction width and doping profile for 8% Sn N-i-P photodiode. e) Schematic showing the*

*evolution of photocarrier diffusion length (dashed arrow) and junction width from low to high Sn content.*

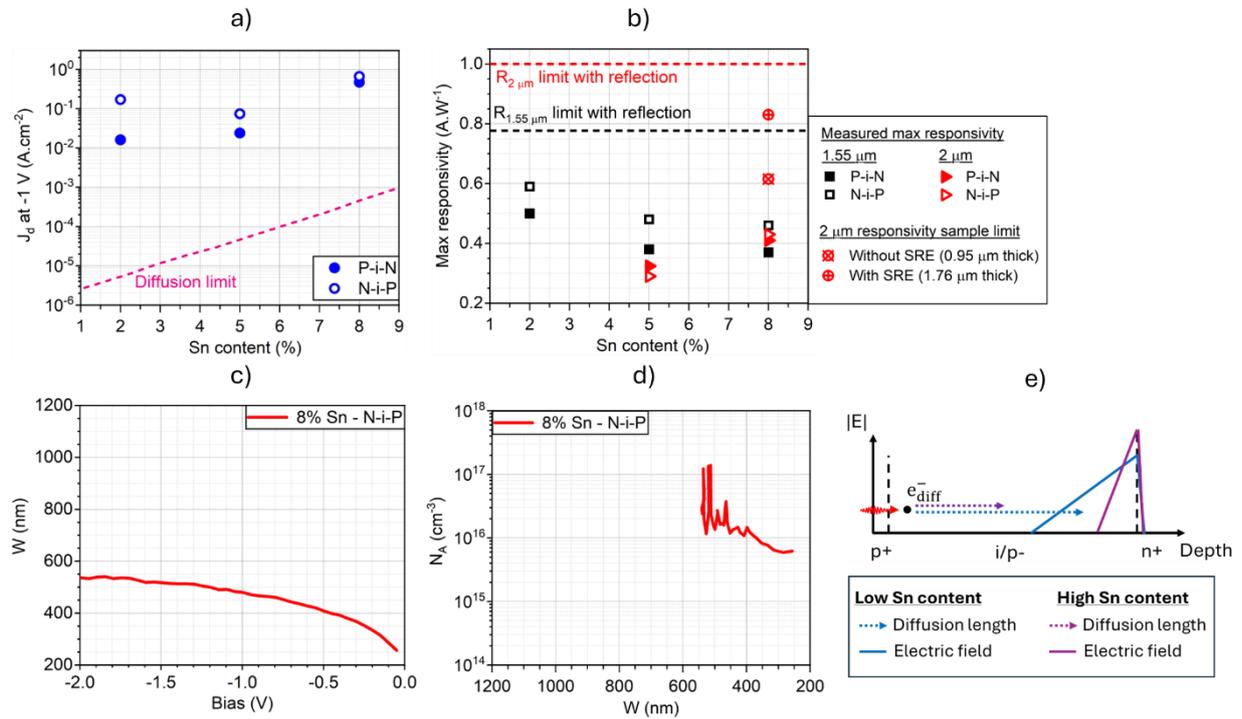

## 2.4 Discussion

### 2.4.1 Summary of P-i-N and N-i-P results

With P-i-N thick absorber GeSn photodiode (2 - 8% Sn), we observed high responsivity on all samples, alongside a significant improvement on detection cutoff wavelength compared to previous results, extending to 2.08 and 2.49 μm at 5% Sn and 8% Sn. The cutoff wavelengths of 2%, 5%, 8% Sn were comparable to commercial SWIR and e-SWIR InGaAs photodiodes. Low dark current density around $2 \times 10^{-2}$ A.cm$^{-2}$ was observed at 2% and 5% Sn P-i-N photodiodes, which we attributed to the junction confinement in high quality GeSn layer, distanced from the top surface by doping the SRE layer in 5% Sn case. Dark current density remained high at 8% Sn P-i-N even with highly doped SRE layer, suggesting a high threading dislocation density in the entire absorber. C-V measurements on P-i-N photodiodes showed junction width significantly shorter than absorber thickness, with p-type background carrier concentration between $10^{15} - 10^{16}$ cm$^{-3}$, suggesting diffusion is the main photocarrier transport mechanism in SWIR range in this case.

By reversing the doping profile from P-i-N to N-i-P, we consistently obtained higher 1.55 µm responsivity on all samples, which was attributed to a switch of photocarrier transport mechanism from diffusion to drift at this wavelength. However, dark current density also increased, which was likely due to the proximity between junction and non-passivated surface, separated only by a thin n+ doped contact layer (40 – 70 nm thick). By comparing P-i-N and N-i-P dark current density and responsivity to their theoretical limit, one can clearly see that there are still big rooms for improvement and highlighted the need for new optimization strategies for thick absorber GeSn photodiodes.

Based on these results, we identified three key elements for the optimization of GeSn thick absorber photodiode performance: top contact layer optimization (material, thickness and doping), compensated doping in the absorber to extend the junction width, and growth quality optimization to extend photocarrier diffusion length and reduce the threading dislocation density. They were discussed in detail in the following section.

## 2.4.2 Optimization strategies for GeSn thick absorber photodiode

Since the passivation efficiency of common dielectric materials like SiO$_2$, SiN or Al$_2$O$_3$ on GeSn remains unclear, one can reduce the impact from surface defects on the dark current density simply by increasing the distance between the surface and the junction, either intrinsically in a P-i-N photodiode (buried junction), or with thicker contact layer (200 nm and above, for example). For N-i-P structure in particular, SWIR transparent window material like Si, Ge or SiGeSn can be selected for thick n+ contact layer, to avoid absorption loss and hence a reduction in responsivity. In addition, the nature of GeSn-metal contact at higher Sn content - Ohmic- or Schottky-like due to Fermi level pinning - and how they depend on the semiconductor doping concentration should be further studied, as the latter contact type required higher field to efficiently collect photocarrier, resulting in a loss of responsivity in PV mode.

As we saw from previous sections, p-type background carrier concentration between $10^{15} - 10^{16}$ cm$^{-3}$ noticeably reduces the junction width compared to the absorber thickness, and as the doping concentration increases with higher Sn content, junction width can become even shorter. Meanwhile, to further increase the responsivity or extend the cutoff wavelength beyond e-SWIR range, absorber thickness needs to be increased up to 3 – 4 µm and/or Sn content needed to be increased beyond 10 %. Diffusion photocurrent, in PV mode in particular, might be significantly reduced in these cases, due to either diffusion length not long enough to transport photocarrier in very thick absorber, or both junction width and diffusion length significantly reduced due to high background carrier concentration and degraded material quality at high Sn content (**Figure 9e**). For example, with background carrier concentration as high as $10^{17}$ cm$^{-3}$ [44,54], depletion width can be below 100 nm for high Sn content GeSn photodiode (10% Sn and above). Meanwhile, recent measurement of carrier lifetime in high Sn content thin GeSn layer reported results around 200-300 ps [74,75], equivalent to a diffusion length around 500 – 600 nm, which can be expected to be even lower for thicker GeSn layer. It should be noted that even N-i-P structure is not exempt from these problems, as long wavelength photons can penetrate deep into the absorber, beyond the junction width and any photocarrier generated there must diffuse back into the junction for collection. To increase the junction width, one can perform compensated n-doping in the absorber. To increase the diffusion length, material quality should be further improved. For such goal, one solution is to continue to improve the Ge buffer quality, via thicker buffer (up to 2.5 µm and beyond, as demonstrated in previous GeSn laser work at 16 - 17% Sn [76,77]) or undoped buffer. More radical solution needs to tackle the lattice mismatch problem between GeSn absorber and Ge buffer, via epitaxial growth of thick graded GeSn layer. It should be noted that current growth strategies for very high Sn content GeSn layer, either via SRE layer or thin graded buffer (both with sacrificial/ graded buffer layer thickness below 500 nm) already embrace this approach, aiming to buckle threading dislocation in the sacrificial/ graded buffer layer and prevent its propagation into the main layer. However, while they work well for thin GeSn layer growth, it remains uncertain whether thin SRE/ buffer layers are capable to maintain similar crystalline quality for significantly thicker layer, beyond 1 μm. Here, past experiences from the heteroepitaxial growth of thick SiGe on Si [78,79], or thick extended InGaAs on InP substrate [65,68,71] can be useful for GeSn photodiode development today. In these works, lattice mismatch grading rate was typically between 0.2 – 0.4%/μm, with total buffer layer thickness ranging between 5 – 20 μm. On GeSn/Ge system, it corresponds to a Sn content grading rate of 1.4 - 2.8%/ μm, before growing the absorber. Finally, novel epitaxial growth scheme like aspect ratio trapping (ART) [51,80,81] can also be explored to geometrically confine threading dislocation via $SiO_2$ sidewalls and thus maintain high crystalline quality in GeSn absorber.

## 3. Conclusion

In this work, we demonstrated a series of thick absorber GeSn photodiodes from 2% Sn to 8% Sn content with high responsivity, low dark current density and high cutoff wavelength. Based on different junction positions in P-i-N and N-i-P doping profile, we identified the contribution of bulk and surface defects on the dark current density, the impact from photocarrier transport mechanism (drift/ diffusion) on the responsivity, and how junction width and photocarrier diffusion length evolved with Sn content. These results helped to identify three key elements for future optimization of GeSn photodiode performance, in particular at high Sn content: thicker and transparent top contact layer to reduce impact from surface defects, compensated doping in the absorber to extend the junction width, and improved material quality to extend the photocarrier diffusion length and reduce threading dislocation density.

**WILEY-VCH**

**Acknowledgements**

The authors acknowledge funding support from Office of Naval Research (ONR) (Grant No. N00014-24-1-2651). J. Rudie acknowledges the funding support from the DoW SMART Scholarship. The authors also acknowledge Lawrence Semiconductor Research Laboratory for the epitaxial growth and SIMS/ XRD characterization of photodiode samples used in this work. This work was performed, in part, at the Center for Integrated Nanotechnologies, an Office of Science User Facility operated for the U.S. Department of Energy (DOE) Office of Science. Los Alamos National Laboratory, an affirmative action equal opportunity employer, is managed by Triad National Security, LLC for the U.S. Department of Energy's NNSA, under contract 89233218CNA000001.

**Conflict of Interest**
The authors declare no conflict of interest.

**Data Availability Statement**
The data that support the findings of this study are available from the corresponding author upon reasonable request.

**Supporting Information**
Supporting Information is available from the Wiley Online Library or from the author.

**References**

[1]   *A Guide to InGaAs and Extended InGaAs Detectors (Brochure - Teledyne Judson)*
[2]   J. Werner, M. Oehme, M. Schmid, M. Kaschel, A. Schirmer, E. Kasper, J. Schulze, *Appl. Phys. Lett.* **2011**, *98*, 061108.
[3]   J. Mathews, R. Roucka, C. Weng, R. Beeler, J. Tolle, J. Menéndéz, J. Kouvetakis, *ECS Trans.* **2010**, *33*, 765.
[4]   H. H. Tseng, H. Li, V. Mashanov, Y. J. Yang, H. H. Cheng, G. E. Chang, R. A. Soref, G. Sun, *Appl. Phys. Lett.* **2013**, *103*, 231907.
[5]   Y.-H. Peng, H. H. Cheng, V. I. Mashanov, G.-E. Chang, *Appl. Phys. Lett.* **2014**, *105*, 231109.
[6]   M. Oehme, K. Kostecki, K. Ye, S. Bechler, K. Ulbricht, M. Schmid, M. Kaschel, M. Gollhofer, R. Körner, W. Zhang, E. Kasper, J. Schulze, *Opt. Express* **2014**, *22*, 839.

[7]     J. Mathews, R. Roucka, J. Xie, S.-Q. Yu, J. Menéndez, J. Kouvetakis, *Appl. Phys. Lett.* **2009**, *95*, 133506.

[8]     R. Roucka, J. Mathews, C. Weng, R. Beeler, J. Tolle, J. Menéndez, J. Kouvetakis, *IEEE J. Quantum Electron.* **2011**, *47*, 213.

[9]     R. T. Beeler, J. Gallagher, C. Xu, L. Jiang, C. L. Senaratne, D. J. Smith, J. Menéndez, A. V. G. Chizmeshya, J. Kouvetakis, *ECS Journal of Solid State Science and Technology* **2013**, *2*, Q172.

[10]    C. Chang, H. Li, C.-T. Ku, S.-G. Yang, H. H. Cheng, J. Hendrickson, R. A. Soref, G. Sun, *Appl. Opt.* **2016**, *55*, 10170.

[11]    B.-J. Huang, J.-H. Lin, H. H. Cheng, G.-E. Chang, *Opt. Lett.* **2018**, *43*, 1215.

[12]    C. Chang, H. Li, S. H. Huang, H. H. Cheng, G. Sun, R. A. Soref, *Appl. Phys. Lett.* **2016**, *108*, 151101.

[13]    Y.-H. Huang, G.-E. Chang, H. Li, H. H. Cheng, *Opt. Lett.* **2017**, *42*, 1652.

[14]    M. Morea, C. E. Brendel, K. Zang, J. Suh, C. S. Fenrich, Y.-C. Huang, H. Chung, Y. Huo, T. I. Kamins, K. C. Saraswat, J. S. Harris, *Appl. Phys. Lett.* **2017**, *110*, 091109.

[15]    S. Su, B. Cheng, C. Xue, W. Wang, Q. Cao, H. Xue, W. Hu, G. Zhang, Y. Zuo, Q. Wang, *Opt. Express* **2011**, *19*, 6400.

[16]    N. Wang, C. Xue, F. Wan, Y. Zhao, G. Xu, Z. Liu, J. Zheng, Y. Zuo, B. Cheng, Q. Wang, *IEEE Photonics J.* **2021**, *13*, 1.

[17]    D. Zhang, C. Xue, B. Cheng, S. Su, Z. Liu, X. Zhang, G. Zhang, C. Li, Q. Wang, *Appl. Phys. Lett.* **2013**, *102*, 141111.

[18]    K.-C. Lee, M.-X. Lin, H. Li, H.-H. Cheng, G. Sun, R. Soref, J. R. Hendrickson, K.-M. Hung, P. Scajev, A. Medvids, *Appl. Phys. Lett.* **2020**, *117*, 012102.

[19]    M. Li, J. Zheng, X. Liu, C. Niu, Y. Zhu, Y. Pang, Z. Liu, Y. Yang, Y. Zuo, B. Cheng, *Opt. Lett.* **2022**, *47*, 4315.

[20]    M. Oehme, M. Schmid, M. Kaschel, M. Gollhofer, D. Widmann, E. Kasper, J. Schulze, *Appl. Phys. Lett.* **2012**, *101*, 141110.

[21]    C.-Y. Chang, R. Bansal, K.-C. Lee, G. Sun, R. Soref, H. H. Cheng, G.-E. Chang, *Opt. Lett.* **2021**, *46*, 3316.

[22]    C.-H. Tsai, K.-C. Lin, C.-Y. Cheng, K.-C. Lee, H. H. Cheng, G.-E. Chang, *Opt. Lett.* **2021**, *46*, 864.

[23]    C.-H. Tsai, B.-J. Huang, R. A. Soref, G. Sun, H. H. Cheng, G.-E. Chang, *Opt. Lett.* **2020**, *45*, 1463.

[24]    R. Bansal, Y.-T. Jheng, K.-C. Lee, S. Wen, Y. Berencén, H.-H. Cheng, G.-E. Chang, *IEEE Journal of Selected Topics in Quantum Electronics* **2025**, *31*, 1.

[25]    X. Li, L. Peng, Z. Liu, Z. Zhou, J. Zheng, C. Xue, Y. Zuo, B. Chen, B. Cheng, *Photonics Res.* **2021**, *9*, 494.

[26]    Y. Dong, W. Wang, D. Lei, X. Gong, Q. Zhou, S. Y. Lee, W. K. Loke, S.-F. Yoon, E. S. Tok, G. Liang, Y.-C. Yeo, *Opt. Express* **2015**, *23*, 18611.

[27]    Q. Huang, J. Zheng, Y. Zhu, X. Liu, Z. Liu, Y. Yang, J. Cui, Z. Liu, Y. Zuo, B. Cheng, *Opt. Lett.* **2024**, *49*, 1365.

[28]    C.-H. Liu, R. Bansal, C.-W. Wu, Y.-T. Jheng, G.-E. Chang, *Adv. Photonics Res.* **2022**, *3*, 2100330.

[29]    J. Zheng, S. Wang, Z. Liu, H. Cong, C. Xue, C. Li, Y. Zuo, B. Cheng, Q. Wang, *Appl. Phys. Lett.* **2016**, *108*, 033503.

[30]    Y. Zhao, N. Wang, K. Yu, X. Zhang, X. Li, J. Zheng, C. Xue, B. Cheng, C. Li, *Chinese Physics B* **2019**, *28*, 128501.

[31]    P.-C. Wang, P.-R. Huang, S. Ghosh, R. Bansal, Y.-T. Jheng, K.-C. Lee, H. H. Cheng, G.-E. Chang, *ACS Photonics* **2024**, *11*, 2659.

[32]    S. Xu, Y.-C. Huang, K. H. Lee, W. Wang, Y. Dong, D. Lei, S. Masudy-Panah, C. S. Tan, X. Gong, Y.-C. Yeo, *Opt. Express* **2018**, *26*, 17312.

[33]  T. Yang, H. Ding, X. Cai, Y. Zhu, J. Qian, J. Yu, G. Lin, W. Huang, S. Chen, C. Li, *IEEE Electron Device Letters* **2024**, *45*, 156.

[34]  M. Oehme, D. Widmann, K. Kostecki, P. Zaumseil, B. Schwartz, M. Gollhofer, R. Koerner, S. Bechler, M. Kittler, E. Kasper, J. Schulze, *Opt. Lett.* **2014**, *39*, 4711.

[35]  H. Wang, J. Zhang, G. Zhang, Y. Chen, Y.-C. Huang, X. Gong, *Opt. Lett.* **2021**, *46*, 2099.

[36]  S. Xu, W. Wang, Y.-C. Huang, Y. Dong, S. Masudy-Panah, H. Wang, X. Gong, Y.-C. Yeo, *Opt. Express* **2019**, *27*, 5798.

[37]  H. Zhou, S. Xu, S. Wu, Y.-C. Huang, P. Zhao, J. Tong, B. Son, X. Guo, D. Zhang, X. Gong, C. S. Tan, *Opt. Express* **2020**, *28*, 34772.

[38]  Q. Chen, H. Zhou, S. Xu, Y.-C. Huang, S. Wu, K. H. Lee, X. Gong, C. S. Tan, *ACS Nano* **2023**, *17*, 12151.

[39]  H. Zhou, S. Xu, Y. Lin, Y.-C. Huang, B. Son, Q. Chen, X. Guo, K. H. Lee, S. C.-K. Goh, X. Gong, C. S. Tan, *Opt. Express* **2020**, *28*, 10280.

[40]  S. Xu, K. Han, Y.-C. Huang, K. H. Lee, Y. Kang, S. Masudy-Panah, Y. Wu, D. Lei, Y. Zhao, H. Wang, C. S. Tan, X. Gong, Y.-C. Yeo, *Opt. Express* **2019**, *27*, 26924.

[41]  H. Cong, C. Xue, J. Zheng, F. Yang, K. Yu, Z. Liu, X. Zhang, B. Cheng, Q. Wang, *IEEE Photonics J.* **2016**, *8*, 1.

[42]  W. Wang, S. Vajandar, S. L. Lim, Y. Dong, V. R. D'Costa, T. Osipowicz, E. S. Tok, Y.-C. Yeo, *J. Appl. Phys.* **2016**, *119*, 155704.

[43]  P. Murkute, N. McKee, D. Plouffe, E. Cho, N. Nooman, T. J. Ronningen, S. Alam, J. Hwang, T. S. Basko, M. Garter, S. Krishna, *Opt. Mater. Express* **2025**, *15*, 2725.

[44]  E. Talamas Simola, V. Kiyek, A. Ballabio, V. Schlykow, J. Frigerio, C. Zucchetti, A. De Iacovo, L. Colace, Y. Yamamoto, G. Capellini, D. Grützmacher, D. Buca, G. Isella, *ACS Photonics* **2021**, *8*, 2166.

[45]  T. Pham, W. Du, H. Tran, J. Margetis, J. Tolle, G. Sun, R. A. Soref, H. A. Naseem, B. Li, S.-Q. Yu, *Opt. Express* **2016**, *24*, 4519.

[46]  H. Tran, T. Pham, J. Margetis, Y. Zhou, W. Dou, P. C. Grant, J. M. Grant, S. Al-Kabi, G. Sun, R. A. Soref, J. Tolle, Y.-H. Zhang, W. Du, B. Li, M. Mortazavi, S.-Q. Yu, *ACS Photonics* **2019**, *6*, 2807.

[47]  M. R. M. Atalla, S. Assali, S. Koelling, A. Attiaoui, O. Moutanabbir, *ACS Photonics* **2022**, *9*, 1425.

[48]  J. Cui, J. Zheng, X. Liu, Y. Wu, J. Li, Q. Huang, Y. Zuo, Z. Liu, B. Cheng, *Appl. Phys. Lett.* **2025**, *126*, 061106.

[49]  C. Cardoux, L. Casiez, E. Kroemer, M. Frauenrath, J. Chrétien, N. Pauc, V. Calvo, J.-M. Hartmann, O. Lartigue, C. Constancias, P. Barritault, N. Coudurier, P. Rodriguez, A. Vandeneynde, P. Grosse, O. Gravrand, A. Chelnokov, V. Reboud, *IEEE Journal of Selected Topics in Quantum Electronics* **2025**, *31*, 1.

[50]  M. Li, J. Zheng, X. Liu, Y. Zhu, C. Niu, Y. Pang, Z. Liu, Y. Zuo, B. Cheng, *Appl. Phys. Lett.* **2022**, *120*, 121103.

[51]  B. Ackland, C. Rafferty, C. King, I. Aberg, J. O'Neill, T. Sriram, A. Lattes, C. Godek, S. Pappas, in *International Image Sensors Workshop*, International Image Sensors Society, **2009**.

[52]  H. Tetzner, W. Seifert, O. Skibitzki, Y. Yamamoto, M. Lisker, M. M. Mirza, I. A. Fischer, D. J. Paul, M. De Seta, G. Capellini, *Appl. Phys. Lett.* **2023**, *122*, DOI: 10.1063/5.0152962.

[53]  A. Giunto, A. Fontcuberta i Morral, *Appl. Phys. Rev.* **2024**, *11*, 041333.

[54]  K. Ye, W. Zhang, M. Oehme, M. Schmid, M. Gollhofer, K. Kostecki, D. Widmann, R. Körner, E. Kasper, J. Schulze, *Solid. State. Electron.* **2015**, *110*, 71.

[55]  N. Rosson, S. Acharya, A. M. Fischer, B. Collier, A. Ali, A. Torabi, W. Du, S.-Q. Yu, R. C. Scott, *Journal of Vacuum Science & Technology B* **2024**, *42*, 052210.

[56] J. Margetis, S.-Q. Yu, N. Bhargava, B. Li, W. Du, J. Tolle, *Semicond. Sci. Technol.* **2017**, *32*, 124006.

[57] W. Dou, M. Benamara, A. Mosleh, J. Margetis, P. Grant, Y. Zhou, S. Al-Kabi, W. Du, J. Tolle, B. Li, M. Mortazavi, S.-Q. Yu, *Sci. Rep.* **2018**, *8*, 5640.

[58] A. Pizzone, S. A. Srinivasan, P. Verheyen, G. Lepage, S. Balakrishnan, J. Van Campenhout, in *2020 IEEE Photonics Conference (IPC)*, **2020**, pp. 1–2.

[59] J. Nguyen, D. Z. Ting, C. J. Hill, A. Soibel, S. A. Keo, S. D. Gunapala, *Infrared Phys. Technol.* **2009**, *52*, 317.

[60] J. Wrobel, E. Plis, W. Gawron, M. Motyka, P. Martyniuk, P. Madejczyk, A. Kowalewski, M. Dyksik, J. Misiewicz, S. Krishna, A. Rogalski, *Sensors and Materials* **2014**, *26*, 235.

[61] S. M. Sze, K. N. Kwok, *Physics of Semiconductor Devices*, Wiley, **2006**.

[62] D. K. Schroder, *Semiconductor Material and Device Characterization*, Wiley, **2005**.

[63] A. Rogalski, *Infrared and Terahertz Detectors*, CRC Press, **2022**.

[64] G. E. Stillman, L. W. Cook, G. E. Bulman, N. Tabatabaie, R. Chin, P. D. Dapkus, *IEEE Trans. Electron Devices* **1982**, *29*, 1355.

[65] R. U. Martinelli, T. J. Zamerowski, P. A. Longeway, *Appl. Phys. Lett.* **1988**, *53*, 989.

[66] M. J. Cohen, G. H. Olsen, in *Proc.SPIE*, **1993**, pp. 436–443.

[67] A. M. Joshi, G. H. Olsen, S. M. Mason, M. Kazakia, V. S. Ban, in *Proc.SPIE*, **1993**, pp. 585–593.

[68] G. H. Olsen, M. J. Lange, M. J. Cohen, D.-S. Kim, S. R. Forrest, in *Proc.SPIE*, **1994**, pp. 151–159.

[69] H. Yuan, G. Apgar, J. Kim, J. Laquindanum, V. Nalavade, P. Beer, J. Kimchi, T. Wong, in *Proc.SPIE*, **2008**, p. 69403C.

[70] Y. Zhang, Y. Gu, Z. Tian, A. Li, X. Zhu, K. Wang, *Infrared Phys. Technol.* **2009**, *52*, 52.

[71] G. H. Olsen, A. M. Joshi, S. M. Mason, K. M. Woodruff, E. Mykietyn, V. S. Ban, M. J. Lange, J. Hladky, G. C. Erickson, G. A. Gasparian, in *Proc.SPIE*, **1990**, pp. 276–282.

[72] G.-E. Chang, S.-Q. Yu, G. Sun, *Sensors* **2023**, *23*, 7386.

[73] H. Tran, W. Du, S. A. Ghetmiri, A. Mosleh, G. Sun, R. A. Soref, J. Margetis, J. Tolle, B. Li, H. A. Naseem, S.-Q. Yu, *J. Appl. Phys.* **2016**, *119*, 103106.

[74] P. Ščajev, V. Soriūtė, G. Kreiza, T. Malinauskas, S. Stanionytė, P. Onufrijevs, A. Medvids, H.-H. Cheng, *J. Appl. Phys.* **2020**, *128*, 115103.

[75] B. Julsgaard, N. von den Driesch, P. Tidemand-Lichtenberg, C. Pedersen, Z. Ikonic, D. Buca, *Photonics Res.* **2020**, *8*, 788.

[76] Q. M. Thai, N. Pauc, J. Aubin, M. Bertrand, J. Chrétien, V. Delaye, A. Chelnokov, J.-M. Hartmann, V. Reboud, V. Calvo, *Opt. Express* **2018**, *26*, 32500.

[77] J. Chretien, Q. M. Thai, M. Frauenrath, L. Casiez, A. Chelnokov, V. Reboud, J.-M. Hartmann, M. El-Kurdi, N. Pauc, V. Calvo, *Appl. Phys. Lett.* **2022**, *120*, 051107.

[78] M. T. Currie, S. B. Samavedam, T. A. Langdo, C. W. Leitz, E. A. Fitzgerald, *Appl. Phys. Lett.* **1998**, *72*, 1718.

[79] Y. Bogumilowicz, J. M. Hartmann, F. Laugier, G. Rolland, T. Billon, N. Cherkashin, A. Claverie, *J. Cryst. Growth* **2005**, *283*, 346.

[80] J.-S. Park, J. Bai, M. Curtin, B. Adekore, M. Carroll, A. Lochtefeld, *Appl. Phys. Lett.* **2007**, *90*, 052113.

[81] H. Stanchu, G. Abernathy, J. Grant, F. M. de Oliveira, Y. I. Mazur, J. Liu, W. Du, B. Li, G. J. Salamo, S.-Q. Yu, *Journal of Vacuum Science & Technology B* **2024**, *42*, 042802.

# Supporting Information

**Systematic study of high-performance GeSn photodiodes with thick absorber for SWIR and extended SWIR detection.**

*Quang Minh Thai, Rajesh Kumar, Abdulla Said Ali, Justin Rudie, Steven Akwabli, Yunsheng Qiu, Mourad Benamara, Hryhorii Stanchu, Kushal Dahal, Xuehuan Ma, Sudip Acharya, Chun-Chieh Chang, Gregory T. Forcherio, Bruce Claflin, Wei Du and Shui-Qing Yu\**

Q. M. Thai, R. Kumar, A. S. Ali, J. Rudie, S. Akwabli, Y. Qiu, S. Acharya, W. Du, S. -Q. Yu
Department of Electrical Engineering and Computer Science, University of Arkansas, Fayetteville, Arkansas 72701, USA
E-mail: syu@uark.edu

A. S. Ali, J. Rudie, K. Dahal, X. Ma, W. Du, S. -Q. Yu
Material Science and Engineering Program, University of Arkansas, Fayetteville, Arkansas 72701, USA

M. Benamara, H. Stanchu, W. Du, S. -Q. Yu
Institute for Nanoscience and Engineering, University of Arkansas, Fayetteville, Arkansas 72701, USA

C. -C. Chang
Center for Integrated Nanotechnologies, Los Alamos National Laboratory, Los Alamos, New Mexico, 87545, USA

G. T. Forcherio
Naval Surface Warfare Center, Electro-Optics Tech. Div. 300 Highway 361, Crane, Indiana 47522, USA

B. Claflin
*Air Force Research Laboratory, Sensors Directorate, Wright-Patterson Air Force Base, Dayton, Ohio 45433, USA*

**SI1. (224) reciprocal maps data and TEM images.**

(224) reciprocal maps for all GeSn photodiode structures were shown in **Figure SI1**. Both Ge and GeSn peak contours were very similar between P-i-N and N-i-P samples of similar Sn content. TEM images of 5% Sn P-i-N and N-i-P photodiodes showed the presence of SRE sacrificial layer on both samples, with half loops formed at the GeSn/Ge interface. (**Figure SI2**).

**Figure SI1.** *(224) reciprocal maps for a) 2% Sn P-i-N, b) 5% Sn P-i-N, c) 8% Sn P-i-N, d) 2% Sn N-i-P, e) 5% Sn N-i-P, f) 8% Sn N-i-P devices. Pseudomorphic lines (with relaxation degree R = 0) and fully relaxation lines (R = 1) were indicated in each plot.*

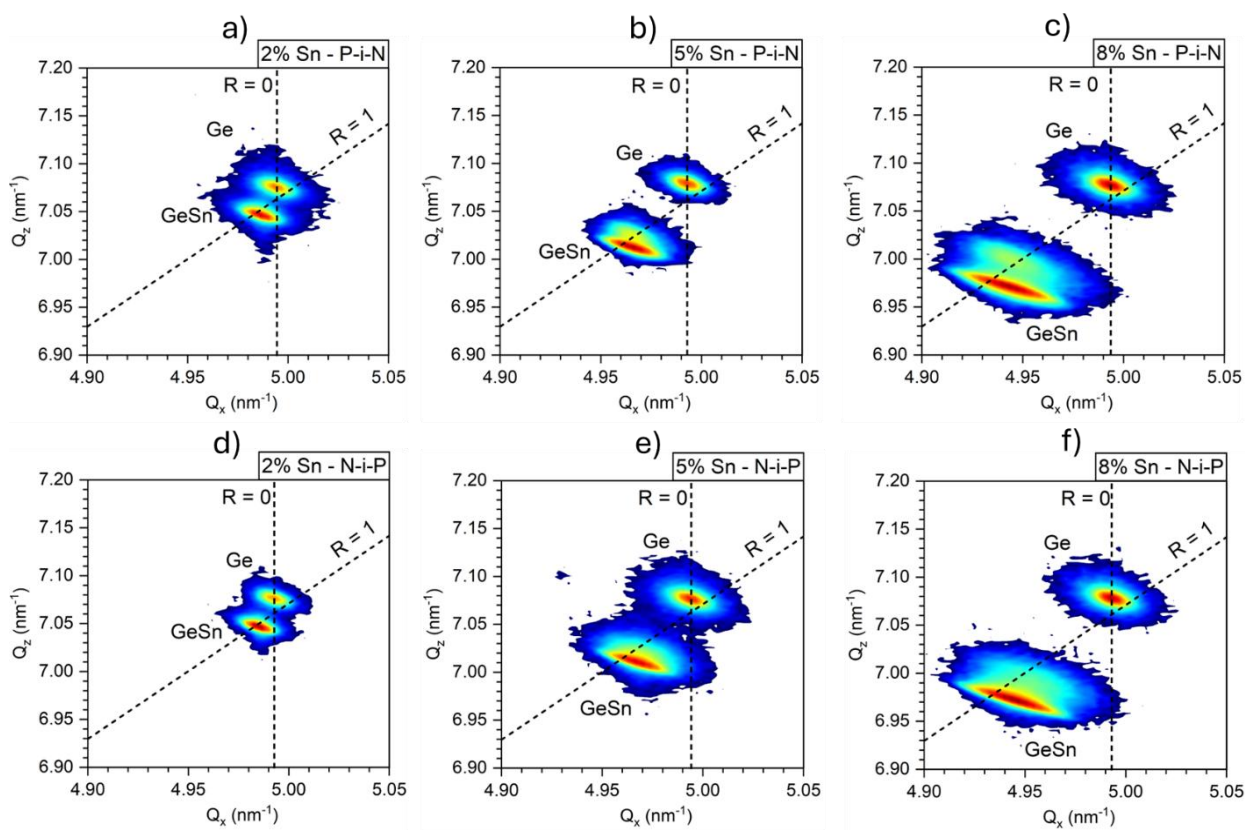

**Figure SI2.** *TEM images of a) 5% Sn P-i-N and b) 5% Sn N-i-P structure, focusing on Ge buffer and GeSn SRE layer.*

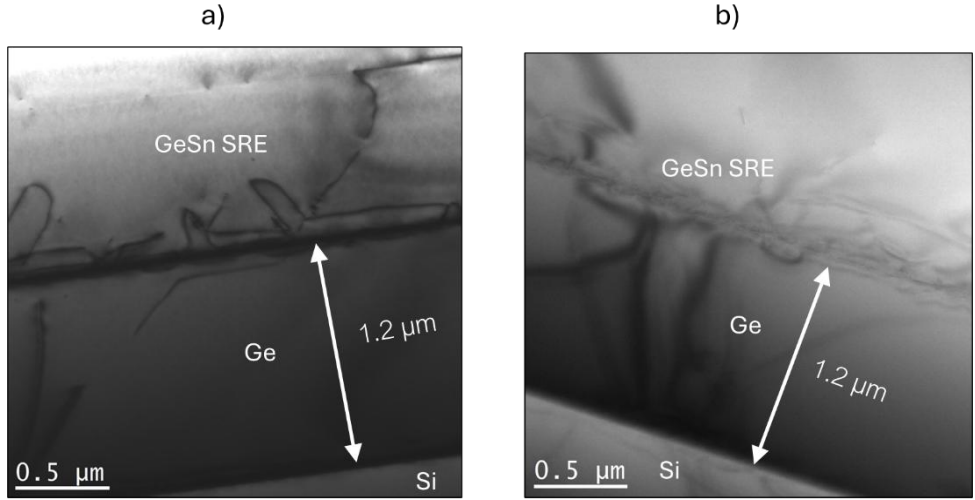

## SI2. Dark current analysis for GeSn P-i-N photodiodes.

The dark current densities of GeSn P-i-N photodiodes (2%, 5% and 8% Sn content) measured from 77 K to 300 K (**Figure SI3**) were used as reference for the dark current analysis that follows.

**Figure SI3.** *Plots of dark current density from 77 K to 300 K for a) 2% Sn , b) 5% Sn and c) 8% Sn P-i-N photodiodes.*

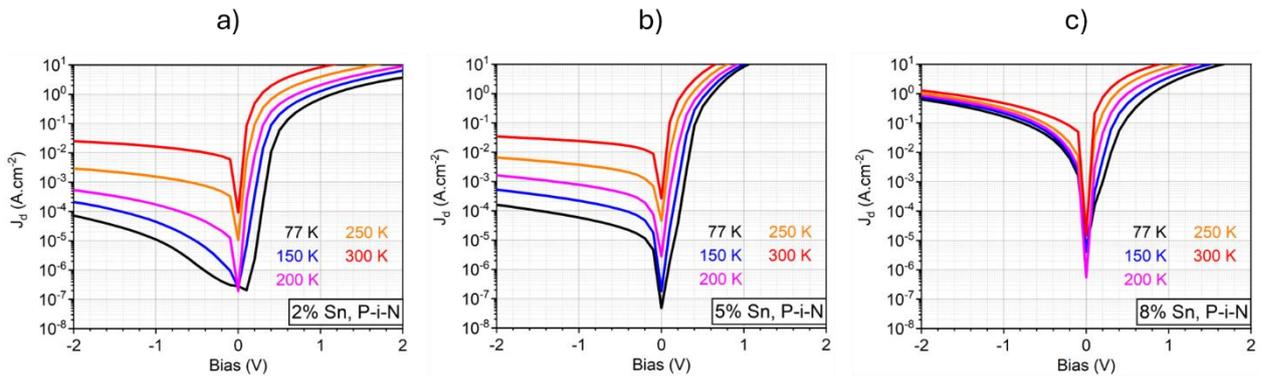

The total simulated dark current density in this work was:

$$J = J_{diff} + J_{GR} + J_{TAT} \qquad (SI1)$$

with $J_{diff}, J_{GR}, J_{TAT}$ the diffusion, SRH generation-recombination (GR), trap-assisted tunneling (TAT) dark current density, respectively. Here, the contribution from leakage current via shunt resistance was neglected. The band-to-band tunneling current density was calculated and found to be extremely small in these cases and thus was also omitted from the simulation. It can be explained by the high effective mass in the exponential term of band-to-band tunneling current density [1], since carriers tunneled between the valence band and the L conduction band in indirect band gap GeSn, both with high effective masses.

$J_{diff}$ was expressed as [1]:

$$J_{diff} = q\left(\frac{D_e}{L_e}\frac{n_i^2}{N_A} + \frac{D_p}{L_p}\frac{n_i^2}{N_D}\right)(e^{\frac{qV}{kT}} - 1) \approx q\left(\frac{D_e}{L_e}\frac{n_i^2}{N_A}\right)(e^{\frac{qV}{kT}} - 1) \qquad (SI2)$$

assuming doping level $N_D$ in the n-contact layer was significantly higher than the p-type background carrier concentration $N_A$ in the absorber, $T$ is the temperature. $J_{GR}$ was expressed as [1]:

$$J_{GR} = -\frac{qn_iW}{2\tau_{GR}} \qquad (SI3)$$

when the reverse bias $|V| > V_{bi}$, with $V_{bi}$ the built-in voltage, taken to be 0.2 V. Finally, $J_{TAT}$ was expressed as [2]:

$$J_{TAT} = \frac{q^2m_t^*M^2N_tV}{8\pi\hbar^3(E_g - E_t)}e^{\frac{-4\sqrt{m_t^*}(E_g-E_t)^{1.5}}{3qE\hbar}} \qquad (SI4)$$

$M^2$ was the trap potential value, taken to be $10^{-23}$ eV$^2$.cm$^3$ [3]. The difference between gap energy and trap energy level $E_g - E_t$ was allowed to be temperature-dependent using the following formula:

$$E_g - E_t(T) = E_0 + \frac{\alpha(T - 300)}{300} \qquad (SI5)$$

A total of six parameters: $L_e$, $\tau_{GR}$, $m_t^*$, $N_t$, $E_0$ and $\alpha$ were optimized in this work, corresponded to the electron diffusion length, GR lifetime, trap effective mass, trap density and trap energy level parameters, respectively. For each sample, fitting was performed using dark current density data at 300 K, 250 K and 150 K as reference. $D_e$, $n_i$, $W$, $E$ were electron diffusion coefficient, intrinsic carrier concentration, depletion width and electric field respectively, and can be further expressed as:

$$D_e(T) = \frac{\mu_e kT}{q} \tag{SI6}$$

$$n_i = \sqrt{N_c N_v} e^{\frac{-E_g}{2kT}} \tag{SI7}$$

$$E = \sqrt{\frac{2qN_A(V_{bi} - V)}{K_r \varepsilon_0}} \tag{SI8}$$

$$W = \sqrt{\frac{2K_r \varepsilon_0 (V_{bi} - V)}{qN_A}} \tag{SI9}$$

with $\mu_e$ the electron mobility (3900 cm$^2$.V$^{-1}$.s$^{-1}$ – taken from Ge value), $N_c$, $N_v$ the conduction and valence band density of states, and $K_r$ the relative permittivity of GeSn ($K_r = 17.6$). As GeSn remained an indirect gap in this Sn content range, $N_c$, $N_v$ values were taken from Ge, since L conduction band effective mass was predicted to be stable regardless of Sn content [4]. The p-type background carrier concentration $N_A$ was set at 4 x 10$^{15}$ cm$^{-3}$, 6 x 10$^{15}$ cm$^{-3}$, 8 x 10$^{15}$ cm$^{-3}$ for 2% Sn, 5% Sn, and 8% Sn P-i-N photodiodes, estimated based on results from C-V measurements (for 8% Sn P-i-N photodiode, such value was estimated based from its N-i-P counterpart results, since we encountered some problems in C-V measurement for this sample as mentioned in the manuscript).

Simulated dark current densities at 300 K, 250 K and 150 K were plotted in **Figures SI4-SI6** for 2%, 5% and 8% Sn P-i-N photodiodes, with fitted parameters values reported in **Table SI1**. For 2% and 5% Sn P-i-N photodiodes, at high temperature, SRH GR was the main contribution to dark current density, while TAT was the main contribution at low temperature (below 150 K), as SRH GR dark current rapidly reduced with temperature (since $n_i$ scaled with $T^{1.5} e^{-E_g/2kT}$, while TAT dark current had a weak dependence to the temperature). Different situation was observed for 8% Sn P-i-N device, where TAT contribution significantly increased and became the dominated factor for all temperatures. The increase of dark current density in 8% Sn P-i-N device compared to 2% and 5% Sn P-i-N devices suggested an increase of trap density $N_t$ and a decrease of GR lifetime $\tau_{GR}$, which increased both TAT and SRH GR dark current density, respectively.

While the simulation matched well with the reference data overall, it should be noted that except for fitted $\tau_{GR}$ values, in particular for 2% Sn and 5% Sn PD where the reference dark current showed a typical $\sqrt{V_{bi} - V}$ dependence at 300 K, all other fitted parameters values should be considered with extreme caution. For $L_e$, since the dark current densities did not show any sign of diffusion limit, with stable values regardless of reverse bias $V$, it was difficult to correctly estimate this parameter. For $m_t^*$, $N_t$, $E_0$ and $\alpha$, they were tangled up in the exponential term of $J_{TAT}$, making it very sensitive to a small change in any of these parameters. Uncertainty from the trap potential value $M^2$ also impacted the fitting accuracy of parameters in TAT term. As a result, multiple sets of solutions for these parameters might exist, and the results for $m_t^*$, $N_t$, $E_0$ and $\alpha$ shown here cannot be used to accurately characterize trap properties in GeSn devices. The reader can notice from Table SI1 that $L_e, m_t^*$, and $E_0$ tended to converge towards the bound limit during the fitting process, and that $N_t$ values were not linearly proportional to $\frac{1}{\tau_{GR}}$, in contrast to prediction from SRH model and emphasized once again the limit of fitted parameters accuracy encountered in this analysis. Other characterization method, like DLTS, should be used instead to obtain a precise estimation of the trap density $N_t$ in these samples.

**Figure SI4.** *Plots of simulated dark current density against the experimental data for 2% Sn P-i-N device, at a) 300K, b) 250 K and c) 150 K.*

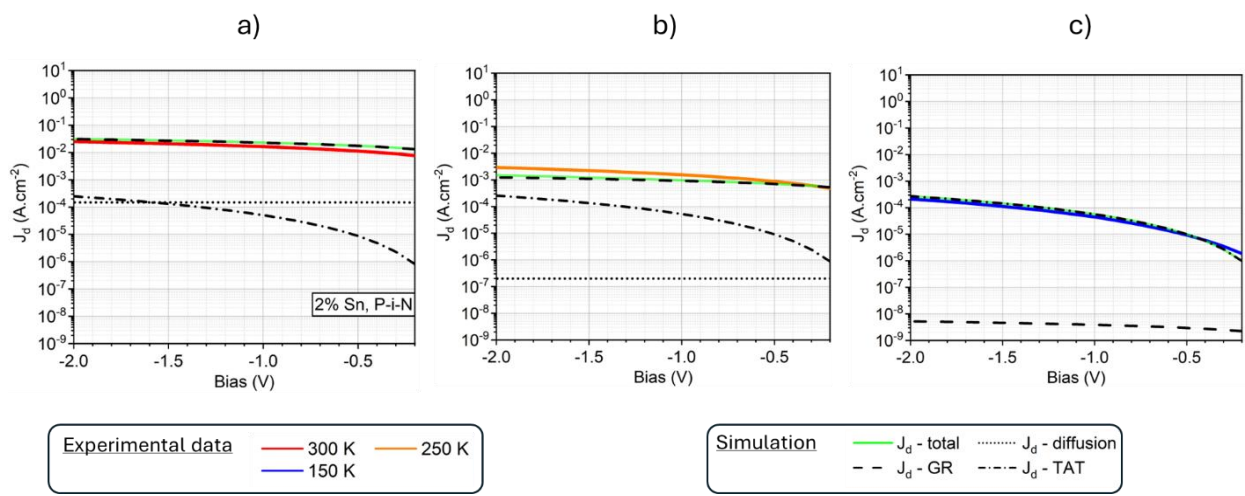

***Figure SI5.*** *Plots of simulated dark current density against the experimental data for 5% Sn*

*P-i-N device, at a) 300K, b) 250 K and c) 150 K.*

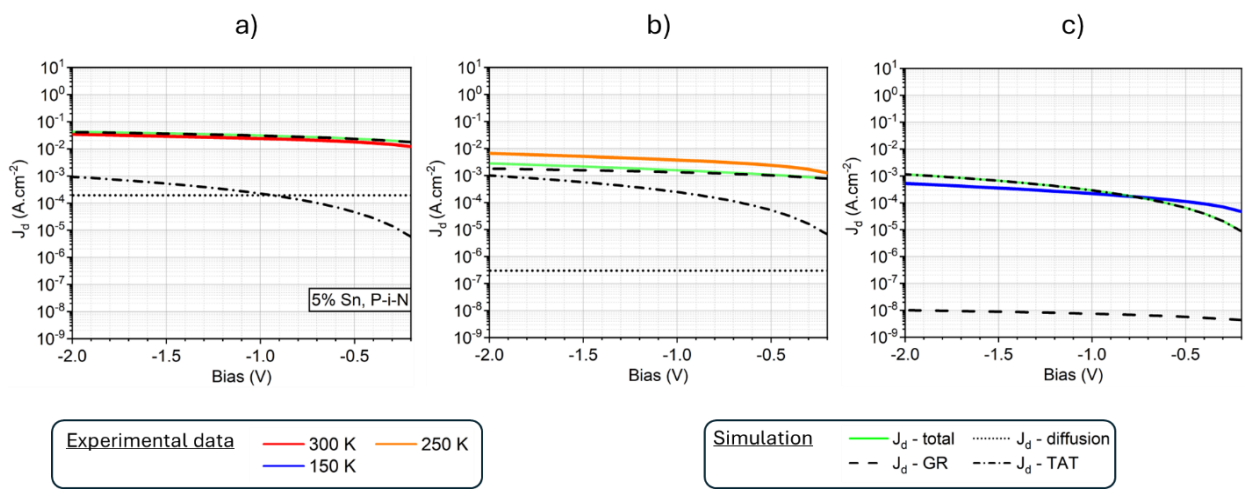

***Figure SI6.*** *Plots of simulated dark current density against the experimental data for 8% Sn*

*P-i-N device, at a) 300K, b) 250 K and c) 150 K.*

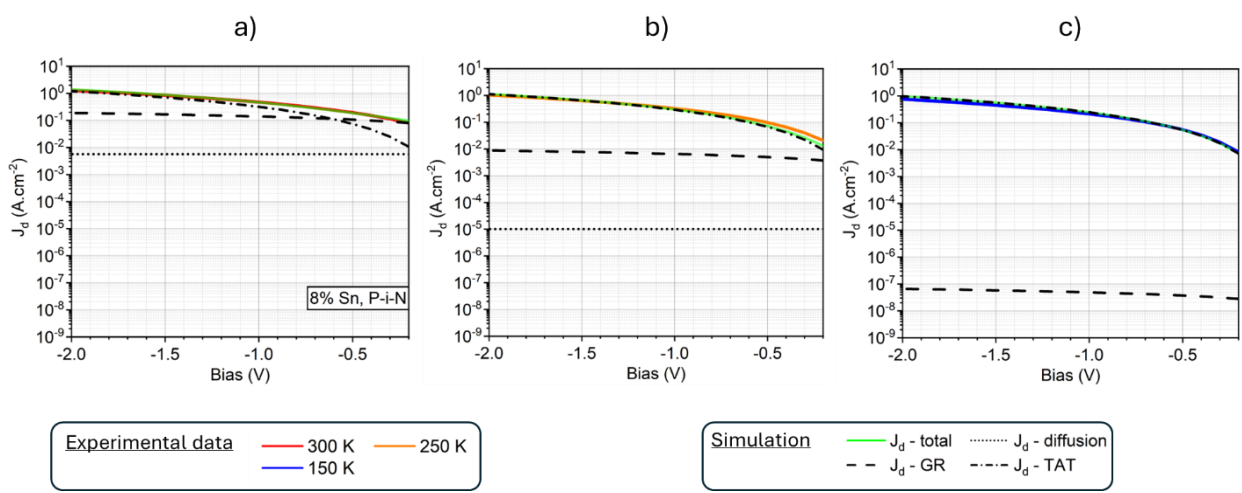

**Table SI1.** *Summary table of fitted parameters used in the dark current analysis $L_e$, $\tau_{GR}$, $m_t^*$, $N_t$, $E_0$ and $\alpha$. ** indicated that the fitted parameter should be considered with extreme caution.*

| Sn content (%) | Design | $\tau_{GR}$ (ns) | $L_e$ (µm) ** | $m_t^*$ ** | $N_t$ (cm$^{-3}$) ** | $E_0$ (eV) ** | $\alpha$ (eV) ** |
|---|---|---|---|---|---|---|---|
| 2 | P-i-N | 7.1 | 200.0 | 0.005 | 2.48 x 10$^{11}$ | 0.1 | 0.004 |
| 5 | P-i-N | 6.2 | 200.0 | 0.005 | 5.85 x 10$^{11}$ | 0.1 | 0.010 |
| 8 | P-i-N | 1.7 | 10.1 | 0.005 | 5.70 x 10$^{14}$ | 0.1 | -0.011 |

## SI3. Other data

Dark current density of 2%, 5% and 8% Sn N-i-P photodiodes were shown in **Figure SI7**. 1.55 and 2 µm responsivities as function of reverse bias for all N-i-P photodiodes were shown in **Figure SI8**. Capacitance, junction widths of 2% and 5% Sn N-i-P photodiodes, alongside the doping profile for 2% Sn N-i-P photodiode were shown in **Figure SI9**. Finally, complete benchmarks of P-i-N and N-i-P GeSn photodiodes performances to other results reported so far, up to 10% Sn content were provided in **Figure SI10**.

**Figure SI7.** *Dark current density from 77 K to 300 K for a) 2% Sn, b) 5% Sn and c) 8% Sn N-i-P photodiodes.*

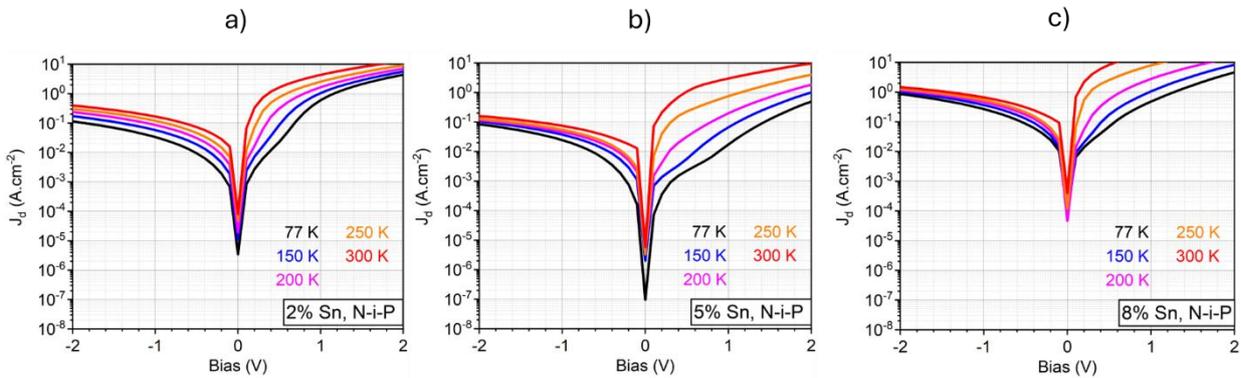

***Figure SI8.*** *1.55 and 2 µm responsivities as function of reverse bias for all N-i-P photodiodes, at 300 K.*

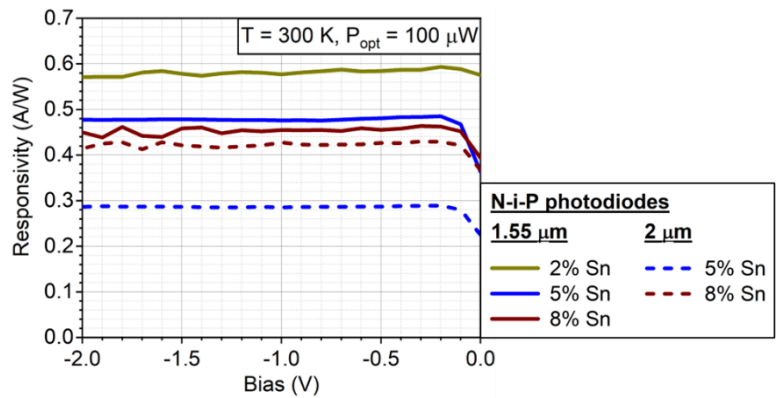

***Figure SI9.*** *a) Capacitance and b) junction width as function of reverse bias for 2% Sn and 5% Sn N-i-P devices. c) Doping concentration as function of junction width for 2% Sn N-i-P device. Doping profile was not calculated for 5% Sn N-i-P devices, since junction width saturated down to very low reverse bias, making it difficult to identify the linear regime and thus correctly calculate the doping profile in the GeSn absorber.*

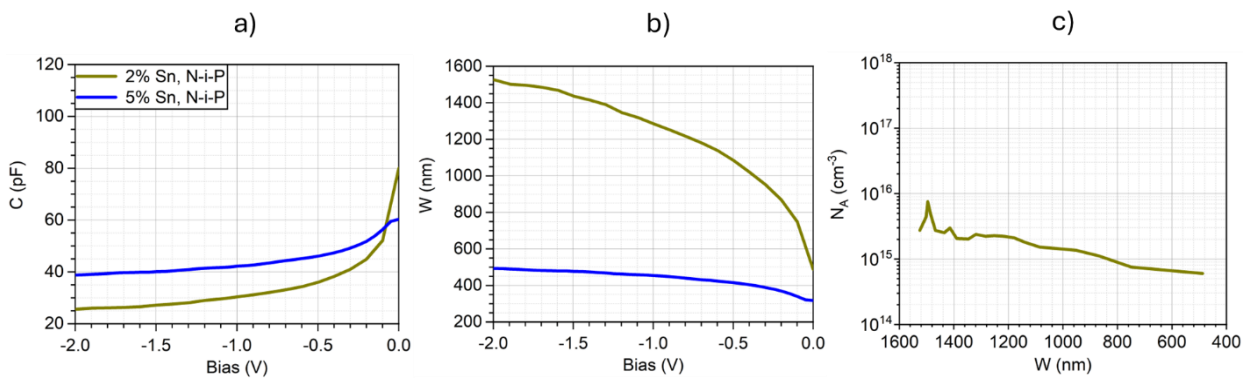

***Figure SI10.*** *Comparison of all P-i-N, N-i-P photodiodes in this work with other GeSn photodiodes results reported in literature. a) Dark current density (top) and normalized dark current density (bottom). b) 1.55 (top) and 2 µm (bottom) maximum responsivity. c) Detection cutoff wavelength (50% of peak intensity). All data were taken at 300 K. For reference numbers, please refer to Figure 6 in the main article.*

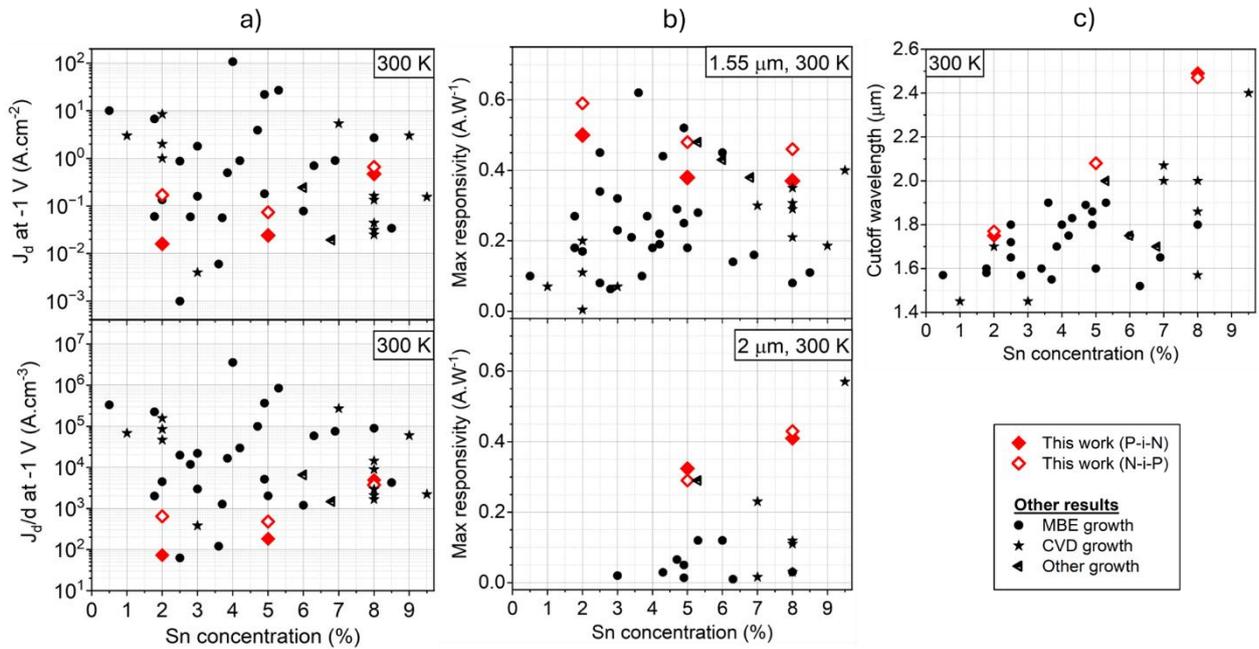

## References

[1] A. Rogalski, *Infrared and Terahertz Detectors*, CRC Press, **2022**.

[2] J. Wrobel, E. Plis, W. Gawron, M. Motyka, P. Martyniuk, P. Madejczyk, A. Kowalewski, M. Dyksik, J. Misiewicz, S. Krishna, A. Rogalski, *Sensors and Materials* **2014**, 26(4), 235

[3] M. A. Kinch, *Journal of Vacuum Science and Technology* **1982**, 21, 215

[4] Q. M. Thai, J. Chretien, M. Bertrand, L. Casiez, A. Chelnokov, V. Reboud, N. Pauc, V. Calvo, *Physical Review B* **2020**, 102 (15), 155203